\documentclass[11pt,tightenlines,eqsecnum,floats,aps,amsmath,amssymb,nofootinbib,prd,shownopacs,floatfix]{revtex4}
\usepackage{graphicx}
\usepackage{epstopdf}
\usepackage{latexsym}
\usepackage{amssymb}
\usepackage{amsmath}
\usepackage{color}
\usepackage{mathrsfs}
\usepackage{xparse}
\usepackage{float}
\usepackage{mathtools}
\usepackage[center]{subfigure}

\begin{document}
  \renewcommand\arraystretch{2}
 \newcommand{\bq}{\begin{equation}}
 \newcommand{\eq}{\end{equation}}
 \newcommand{\bqn}{\begin{eqnarray}}
 \newcommand{\eqn}{\end{eqnarray}}
 \newcommand{\nb}{\nonumber}
 \newcommand{\lb}{\label}
 \newcommand{\cb}{\color{blue}}
    \newcommand{\cc}{\color{cyan}}
        \newcommand{\cm}{\color{magenta}}
\newcommand{\rc}{\rho^{\scriptscriptstyle{\mathrm{I}}}_c}
\newcommand{\rd}{\rho^{\scriptscriptstyle{\mathrm{II}}}_c} 
\NewDocumentCommand{\evalat}{sO{\big}mm}{%
  \IfBooleanTF{#1}
   {\mleft. #3 \mright|_{#4}}
   {#3#2|_{#4}}%
}
\newcommand{\PRL}{Phys. Rev. Lett.}
\newcommand{\PL}{Phys. Lett.}
\newcommand{\PR}{Phys. Rev.}
\newcommand{\CQG}{Class. Quantum Grav.}
\newcommand{\parallelsum}{\mathbin{\!/\mkern-5mu/\!}}

\title{Quantum gravitational onset of Starobinsky inflation in a closed universe}

\author{Lucia Gordon$^{1}$}
\email{luciagordon@college.harvard.edu}
\author{Bao-Fei Li $^{2}$}
\email{baofeili1@lsu.edu}
\author{Parampreet Singh$^2$}
\email{psingh@lsu.edu}
\affiliation{$^{1}$ Department of Physics, Harvard University,
Cambridge, MA 02138, USA \\
$^{2}$ Department of Physics and Astronomy, Louisiana State University, Baton Rouge, LA 70803, USA}

\begin{abstract}
Recent cosmic microwave background observations favor low energy scale inflationary models in a closed universe. However, onset of inflation in such models for a closed universe is known to be severely problematic. In particular, such a universe recollapses within a few Planck seconds and encounters a big crunch singularity when initial conditions are given in the Planck regime. We show that this problem of onset of inflation in low energy scale inflationary models can be successfully overcome in a quantum gravitational framework where the big bang/big crunch singularities are resolved and a non-singular cyclic evolution exists prior to inflation. As an example we consider a model in loop quantum cosmology and demonstrate that the successful onset of low energy scale inflation in a closed universe is possible for the Starobinsky inflationary model starting from a variety of initial conditions where it is impossible in the classical theory. For comparison we also investigate the onset of inflation in the $\phi^2$ inflationary model under highly unfavorable conditions and find similar results. Our numerical investigation including the phase space analysis shows that the pre-inflationary phase with quantum gravity effects is composed of non-identical cycles of bounces and recollapses resulting in a hysteresis-like phenomenon, which plays an important role in creating suitable conditions for inflation to occur after some number of non-singular cycles. Our analysis shows that the tension in the classical theory amounting to the unsuitability of closed FLRW universes with respect to the onset of low energy scale inflation can be successfully resolved in loop quantum cosmology. 
\end{abstract}

\maketitle

\section{Introduction}
\label{introduction}
\renewcommand{\theequation}{1.\arabic{equation}}\setcounter{equation}{0}

Cosmic microwave background (CMB) observations suggest that inflationary models with an inflaton in a plateau potential have been favored since Planck 2013 \cite{Planck2013,Planck2015, Planck2018}. The inflationary models that are consistent  with the Planck constraints on the spectral index and the tensor-to-scalar ratio for the pivot mode include the Starobinsky model \cite{Starobinsky}, the Higgs inflationary model \cite{sbb1989, bs2008} and a broad class of cosmological attractor models \cite{klr2013,ck2014,gklr2015,klr2014}. In these models, inflation takes place at an energy scale far below the Planck density. For this reason, they are also called the low energy scale inflationary models. But if the universe is closed, then the onset of inflation in such models is severely problematic \cite{lindelecture, linde2018,ckl2015}. This problem, which is tied to the recollapse of the universe and the big crunch singularity, can be avoided in spatially-flat and open universes \cite{lindelecture, linde2018}. However, the recent Planck Legacy 2018 (PL 2018) release has confirmed the presence of an enhanced lensing amplitude in the CMB power spectra  \cite{planckcp, planckpower}, which favors a closed universe over a spatially-flat one \cite{nature}. Consequently, a pertinent question is the following: How does a closed universe starting from the Planck regime lead to successful low scale inflation? To answer this question, one needs a framework which goes beyond the classical description of spacetime and allows a robust resolution of the big crunch singularity while alleviating the problem of initial conditions for low scale inflation in a closed universe.

While the low energy inflationary model in a closed universe starting from the Planck regime typically ends in a big crunch singularity in a few Planck seconds,  
the onset of inflation in a closed universe is not as problematic for the chaotic inflationary models such as $\phi^2$ inflation which, however, is not favored by the Planck data. In these models, where inflation can occur at higher energy scales, the initial conditions of the inflaton starting from a single Planck sized domain can be such that the kinetic and gradient energy of the scalar field are smaller than its potential energy which is order $U\sim1$ (in Planck units). Since the potential energy is dominant, the universe can  avoid the recollapse and the subsequent big crunch singularity, and the dynamical evolution successfully results in a phase of inflation shortly after the initial time. Note that this is true only for those initial conditions where potential energy is dominant. Of course, if the inflaton starts with kinetic energy domination then the fate of such a universe even in a $\phi^2$ inflationary model can be similar to that of low energy scale inflation in the sense that the universe encounters a big crunch singularity before a phase of inflation can start. Previous studies show that the probability of the quantum creation of a closed universe in which inflation takes place at the energy scale $U\ll 1$ is exponentially suppressed \cite{linde1984a,linde1984b,v1984,v1988,v1989}. 
However, all of the above arguments exclude the role of non-perturbative quantum gravity effects in the Planck regime. Further, an important problem irrespective of the spatial curvature of the universe is that inflationary models are past-incomplete in the classical theory \cite{BV94,BGV}. Since quantum gravity effects are expected to resolve spacetime singularities, it is quite possible that the above situation changes dramatically when one considers a quantum gravitational version of the model. 

The goal of this paper is to show that the above problem of the onset of inflation in a closed universe can be successfully resolved in a setting motivated by loop quantum gravity where big bang and big crunch singularities are replaced by a big bounce due to quantum geometric effects. In our analysis we consider the Starobinsky inflationary model as an example of a low energy inflationary model and study the onset of inflation for a variety of initial conditions. For comparison we also investigate the $\phi^2$ model for initial conditions which are not favorable in the classical theory for the onset of inflation. While our analysis uses techniques of loop quantum cosmology (LQC) for the chaotic and Starobinsky potentials, the essential idea and results can be replicated for any bouncing model and other inflationary potentials.

Our framework is based on a loop quantization of a closed universe performed in LQC which is 
 a non-perturbative quantization of cosmological models based on loop quantum gravity  \cite{review2}. Unlike the Wheeler-DeWitt quantum cosmology where the spacetime manifold is differentiable, LQC is based on a discrete quantum geometry predicted by loop quantum gravity. The quantum evolution is dictated by a non-singular quantum difference equation which results in a resolution of the big bang singularity by replacing it with a quantum bounce when the spacetime curvature becomes Planckian \cite{aps1,aps3,slqc}, including in the presence of inflationary potentials \cite{gls,aps4}, anisotropies and inhomogenities \cite{review2}. Using the consistent histories formalism one can also compute the probability of the  bounce which turns out to be unity \cite{craig-singh}. %
  For a closed universe, loop quantum gravitational effects resolve both the past and future singularities, resulting in a cyclic universe \cite{apsv,warsaw,ck2011}. It turns out that the quantum evolution can be very well approximated by the effective dynamics obtained from an effective spacetime description \cite{aps3,apsv,numlsu-2,numlsu-4}. 
  Using this description singularity resolution has been  established to be a generic feature for various isotropic and anisotropic models in LQC \cite{ps09,ps12}, and various phenomenologically interesting consequences for the early universe and resulting signatures in the CMB have been discussed \cite{as2017}.  For the spatially-flat model it was found that the inflationary dynamics is an attractor after the bounce for $\phi^2$, power-law inflation and the Starobinsky potential \cite{svv,ranken-ps,gs-2013,lsw2018b,tao2017,bg2016,bg2016b,shahalama,shahalamb},  and the probability for inflation to occur is very large \cite{as2011,ck2011b,cs2014}. For closed universes, the dynamics with the $\phi^2$ inflationary potential in the presence of spatial curvature has been previously studied in LQC \cite{ds2020},  which revealed some novel features in comparison to the spatially-flat case. It was found that due to the recollapse caused by the spatial curvature and the bounce caused by quantum geometry, the evolution of a closed universe filled with a homogeneous scalar field in an inflationary potential is usually characterized by a number of cycles in the pre-inflationary era with a hysteresis-like phenomenon due to the asymmetry in the equation of state (or equivalently the asymmetry in the pressure of the scalar field) in each cycle. It was found that quantum geometric modifications in LQC enhance this hysteresis-like phenomenon in comparison to previously studied bouncing models where this phenomenon was noted earlier \cite{kanekar,st2012,sst2015}. Owing to this hysteresis-like behavior, even when starting from unfavorable initial conditions for inflation, the universe comes out of non-inflationary cyclic evolution and enters into a phase of inflation. Once the universe is in this phase, the scalar field starts to play a dominant role and the effects of the spatial curvature become negligible due to the exponential expansion of the universe, implying that no further cycles occur. Though the work in Ref. \cite{ds2020} found evidence of inflation occurring after hysteresis in $\phi^2$ inflation, the robustness of the existence of the inflationary phase and its possible generalization to low energy scale inflation models was not studied.

The aim of this paper is to reconcile the closed universe scenario with low scale inflation in the framework of LQC for the Starobinsky potential.\footnote{Unlike the conventional formalism where the Starobinsky potential arises from a  higher order correction in the action, in LQC the Starobinsky potential is taken as a phenomenological input. For higher order actions in LQC, see \cite{olmo-ps}.} Our strategy follows the encouraging results for $\phi^2$ inflation in the closed model of LQC \cite{ds2020} with an aim to understand the onset of inflation with the Starobinsky potential and also to gain further insights by comparing it with $\phi^2$ inflation. Using the effective Hamilton's equations resulting from the holonomy quantization for a closed LQC universe \cite{apsv}, we consider a single scalar field minimally coupled to gravity with the initial conditions set at the maximum energy density at some initial volume. We explore the background evolution of the universe starting with different types of initial conditions, with varying ratios of kinetic and potential energy, some of which are unfavorable for inflation to start in the classical theory, and perform numerical simulations for the  $\phi^2$ and Starobinsky potentials up through the onset of inflation. To investigate some aspects of the qualitative behavior we also study two-dimensional phase space portraits and find inflationary separatrices and other cosmological attractors in the plots. Our results show that both $\phi^2$ inflation and Starobinsky inflation can take place under a variety of initial conditions starting from the Planck regime in the closed model. The primary reason for this is the hysteresis-like phenomena which  are  found to be much weaker in Starobinsky inflation. Due to this, the onset of inflation is delayed in comparison to $\phi^2$ inflation.

This manuscript is organized as follows. In Sec. \ref{review}, we briefly review the effective dynamics of a closed FLRW universe using the holonomy quantization in LQC. We obtain the necessary equations which include the Hamilton's equations and the modified Friedmann equation  for the purpose of a detailed analysis of the background dynamics of the universe when gravity is minimally coupled to a single scalar field with an inflationary potential. In Sec. \ref{chaotic}, we consider $\phi^2$ inflation in a closed FLRW universe in LQC with the initial conditions given in the Planck regime. We address the general properties of the pre-inflationary dynamics and present phase space portraits with trajectories starting from various  initial conditions that all end up with a reheating phase, corresponding to the spiral at the center of the plots. In Sec. \ref{starobinsky},  we study the Starobinsky potential in a closed universe in LQC. We show that after taking into account quantum gravitational effects, many initial conditions that are unfavorable for inflation in   classical cosmology do in fact lead to inflation at late times in LQC. The phase space portraits will be presented in order to show the qualitative behavior of numerical solutions starting with various initial conditions. Finally, in Sec. \ref{summary}, we summarize the main results concerning $\phi^2$ and Starobinsky inflation in a closed universe in LQC and discuss the differences and similarities between these two inflationary models.

In our paper, we use Planck units with $\hbar=c=1$ while keeping Newton's constant $G$ explicit in our formulas. In the numerical analysis, $G$ is also set to unity. 

\section{Effective dynamics of a closed universe in Loop Quantum Cosmology}
\label{review}
\renewcommand{\theequation}{2.\arabic{equation}}\setcounter{equation}{0}

In LQC,  the quantization of the homogeneous and isotropic FLRW universe is carried out in terms of the Ashtekar-Barbero connection $A^i_a$ and its conjugate triad $E^i_a$. The non-perturbative modifications to the classical Hamiltonian constraint arise from the regularization of the field strength  of the connection and the inverse volume terms. It turns out that the latter are not dominant compared to the field strength modifications for singularity resolution \cite{apsv} and quickly decay for volumes greater than the Planck volume. For this reason we will only focus on the effective dynamics incorporating modifications to the field strength of the connection. In the literature, the regularization of the field strength in a closed FLRW universe has been explored in two different  ways. One is based on the quantization of the holonomy of the connection over closed loops \cite{apsv}, the other is the connection based quantization \cite{ck2011}. Both of these  approaches, which can be viewed as quantization ambiguities, lead to singularity resolution and give similar qualitative behavior away from the bounce regime \cite{ds2017}. For the $\phi^2$ potential a comparative analysis of these two quantizations was performed in Ref. \cite{ds2020}, which revealed 
the robustness of the  qualitative features of hysteresis and the subsequent inflationary phase. In the following, we will study the holonomy based quantization and we expect the main results to  hold  for the connection based quantization as well.

In a  cosmological setting, due to the  homogeneity and isotropy of the universe, the Ashtekar-Barbero connection and its conjugate triad can be  symmetry-reduced to the canonical pair of $c$ and $p$ \cite{apsv}. This set of canonical variables is equivalent to a new set of variables, namely $b$ and $v$ that are  commonly used in the $\bar \mu $ scheme \cite{aps3}.  In a closed universe, the physical volume of the unit sphere  spatial manifold is given by $v=|p|^{3/2}=2\pi^2a^3$ with $a$ representing the scale factor of the universe. The  conjugate variable $b$ is defined via $b=c|p|^{-1/2}$. Besides the gravitational degrees of freedom, in order to initiate the onset of  inflation,  one also needs degrees of freedom in the matter sector which are the scalar field  $\phi$ and its conjugate momentum $p_\phi$. These  fundamental canonical pairs in the phase space satisfy the  Poisson brackets: 
\bq
\{b,v\}=4\pi G\gamma, ~~~~ \{\phi, p_\phi\}=1,
\eq
where $\gamma$ is the Barbero-Immirzi parameter fixed by black hole thermodynamics in LQG. As is usual in LQC, we take the value of $\gamma \approx 0.2375$. 

In terms of the canonical variables introduced above, the effective Hamiltonian constraint of a closed FLRW universe for the holonomy quantization takes the form \cite{apsv}
\bq
\lb{hamiltonian}
\mathcal H_\mathrm{eff}=-\frac{3v}{8\pi G \gamma^2\lambda^2}\Big[\sin^2(\lambda b-D)-\sin^2D+(1+\gamma^2)D^2\Big]+\mathcal H_\mathrm{m},
\eq
where $D$ is defined by
\bq
\lb{2.1}
D=\lambda\left(\frac{2\pi^2}{v}\right)^{1/3},
\eq 
and $\lambda$($=2\sqrt{\sqrt{3}\pi\gamma}$) is  the minimum area eigenvalue in LQG. Since we only consider a single massive scalar field coupled to gravity, the matter sector of the Hamiltonian constraint denoted by $\mathcal H_\mathrm m$ is given  by
\bq
\lb{2.2}
\mathcal H_\mathrm m= \frac{p_\phi^2}{2v}+v \, U,
\eq 
where $U$ refers to the potential of the scalar field. From the total Hamiltonian constraint (\ref{hamiltonian}), one can derive in a straightforward way the equations of motion  for each canonical variable, which turn out to be 
\bqn
\lb{vdot}
\dot v&=& \frac{3v}{\lambda \gamma}\sin\left(\lambda b -D\right)\cos\left(\lambda b-D\right),\\
\lb{bdot}
\dot b &=&-4\pi G \gamma \Big[\rho-\rho_1+P\Big],\\
\lb{phidot}
\dot \phi&=&\frac{p_\phi}{v}, \quad \quad \dot p_\phi=-v \, U_{,\phi},
\eqn
here $U_{,\phi}$ denotes the derivative of the potential with respect to the scalar field. The energy density $\rho$ and the pressure $P$ are defined  respectively by $\rho=\mathcal H_m/v$ and $P=-\partial \mathcal H_m/\partial v$. In terms of the scalar field and its momentum, they are explicitly given by 
\bq
\lb{2.4}
\rho=\frac{p^2_\phi}{2v^2}+U, \quad \quad P=\frac{p^2_\phi}{2v^2}-U.
\eq
Moreover, in  Eq. (\ref{bdot}), $\rho_1$ is given by 
\bq
\lb{2.5}
\rho_1=\frac{D \rho_\mathrm {crit}}{3}\Big[2(1+\gamma^2)D-\sin\left(2\lambda b-2D\right)-\sin\left(2D\right)\Big],
\eq
here $\rho_\mathrm{crit} = 3/8\pi G\gamma^2\lambda^2$. It turns out that this is the maximum energy density allowed in a spatially-flat FLRW universe by LQC where the bounce occurs. In the spatially closed model, the bounce density can be different. From Eq. (\ref{vdot}) and the vanishing of the total Hamiltonian constraint, it is straightforward to find the modified Friedmann equation which is
\bq
\lb{friedmann}
H^2=\frac{\dot v^2}{9v^2}=\frac{8\pi G}{3}\left(\rho-\rho_\mathrm{min}\right)\left(1-\frac{\rho-\rho_\mathrm{min}}{\rho_\mathrm{crit}}\right),
\eq
where 
\bq
\lb{low}
\rho_\mathrm{min}=\rho_\mathrm{crit}\Big[\left(1+\gamma^2\right)D^2-\sin^2D\Big].
\eq
Since the right-hand side of the Friedmann equation is non-negative, the energy density at any moment during the  evolution has to satisfy the condition 
\bq
\lb{2.6}
\rho_\mathrm{min}\le \rho\le \rho_\mathrm{max},
\eq
where $\rho_\mathrm{max}$ is defined by 
\bq 
\rho_\mathrm{max}=\rho_\mathrm{min}+\rho_\mathrm{crit}.
\eq
It should be noted that unlike $\rho_\mathrm{crit}$, the maximum and minimum energy densities in a closed universe depend explicitly on the volume of the universe and thus do not have fixed values. The bounces and recollapses of a closed universe happen at the turning points when the Hubble rate vanishes, which is equivalent to the condition $\rho=\rho_\mathrm{min}$ or $\rho=\rho_\mathrm{max}$. The character of the turning point, namely whether it is a bounce point or a recollapse point, depends on the second derivative of the volume. To be specific, a turning point is a bounce point when $\ddot v>0$ and a recollapse point when $\ddot v<0$. From the Hamilton's equations (\ref{vdot})-(\ref{bdot}), it is straightforward to find that when $\rho=\rho_\mathrm{min}$
\bq
\ddot v|_{\rho_\mathrm{min}}=-12\pi G v \left(\rho+P-\rho_2\right),
\eq
and when $\rho=\rho_\mathrm{max}$ 
\bq
\ddot v|_{\rho_\mathrm{max}}=12\pi G v \left(\rho+P-\rho_2\right),
\eq
where
\bq\label{rho2}
\rho_2=\frac{D\rho_\mathrm{crit}}{3}\Big[2(1+\gamma^2)D-\sin(2D)\Big].
\eq
As a result, depending on the initial conditions, the bounces and recollapses can occur at either the maximum or the minimum energy density. Furthermore, at the turning point when the Hubble rate vanishes, from  the equation of motion (\ref{vdot}), one can find $\sin(2\lambda b-2D)$ also vanishes. Therefore, $\rho_1=\rho_2$ at a bounce or recollapse point which indicates that  $\ddot v$  and $\dot b$ have the same sign  at the minimum energy density and opposite signs at the maximum energy density.

In addition, it can be shown in a straightforward way that the modified Friedmann equation has the right classical limit.  In Eq. (\ref{low}), if we take the limit $v\gg1$, then $\rho_\mathrm{min}\approx \rho_\mathrm{crit}\gamma^2 D^2=\frac{3}{8\pi G a^2}$ and  the Friedmann equation (\ref{friedmann}) reduces to 
\bq
\lb{2.7}
H^2=\frac{8\pi G}{3}\left(\rho-\frac{\rho^2}{\rho_\mathrm{crit}}+2\rho \gamma^2 D^2-\rho_\mathrm{crit}\gamma^2 D^2-\rho_\mathrm{crit}\gamma^4 D^4\right).
\eq
Considering the classical limit $\rho\ll1$ and $a\gg 1$, the above equation can be further reduced to 
\bqn
\lb{2.8}
H^2&=&\frac{8\pi G}{3}\rho-\frac{8\pi G}{3}\gamma^2 D^2 \rho_\mathrm{crit},\nb \\
&=&\frac{8\pi G}{3}\rho-\left(\frac{2\pi^2}{v}\right)^{2/3}=\frac{8\pi G}{3}\rho-\frac{1}{a^2},
\eqn
which is exactly the classical Friedmann equation for a closed universe.  Since  loop quantization is only applied to the geometrical sector of the classical phase space, the equations of motion in the matter sector are not changed by the quantum geometrical effects.  By using the Hamilton's equations of the scalar field in (\ref{phidot}), it is straightforward to show that the Klein-Gordon equation and thus the continuity equation also hold in a closed universe of LQC. 

Finally, we would like to briefly review the hysteresis-like phenomenon in a cyclic universe filled with a homogeneous scalar field \cite{st2012,ds2020}. Assuming the evolution of the cyclic universe is adiabatic,  the work done by the scalar field during each contraction-expansion cycle can be explicitly computed as 
\bq
W=\oint Pdv=\int_\mathrm{contraction} Pdv +\int_\mathrm{expansion} Pdv.
\eq
On the other hand, the change in the total energy of the scalar field at two consecutive recollapse points is simply given by 
\bq
\delta M=\rho^{(i)}_\mathrm{rec}v^{(i)}_\mathrm{rec}-\rho^{(i-1)}_\mathrm{rec}v^{(i-1)}_\mathrm{rec},
\eq
where $\rho^{(i)}_\mathrm{rec}$/$v^{(i)}_\mathrm{rec}$ denotes the energy density/volume at the $i$th recollapse point. Using the energy conservation law $W+\delta M=0$, one can relate the change in the maximum volume of the universe at two successive turnarounds to the net work done by the scalar field in one complete cycle. This relationship in general is model-dependent and also determined by the character of the turnarounds in each model. In the current case with the effective dynamics determined by the modified Friedmann equation (\ref{friedmann}), assuming the energy densities at the recollapse points are given by the minimum density $\rho_\mathrm{min}$, then when $v^{(i)}$ is much greater than unity,
the difference in the volumes of two consecutive recollapses can easily be shown to be  
\bq
\lb{changeinvolume}
\delta v^{1/3}_\mathrm{rec}=\frac{-\oint P dv}{(2\pi^2)^{2/3}\rho_\mathrm{crit}\gamma^2\lambda^2}.
\eq
As a result, there is a  change in the maximum volume in each cycle which is directly related to the asymmetry of the pressure of  the scalar field during the  contraction and expansion phases of each cycle. This results in hysteresis-like behavior which becomes evident via the plots of the  equation of state \cite{st2012,ds2020}. We find that an increase in the maximum volume and a decrease in the equation of state in each cycle will play an important role in the onset of inflation even with initial conditions which are unfavorable in the classical theory.

\section{$\phi^2$ inflation in a closed FLRW universe in LQC}
\lb{chaotic}
\renewcommand{\theequation}{3.\arabic{equation}}\setcounter{equation}{0}
Before we discuss the Starobinsky potential case in the next section, we study the occurrence of $\phi^2$ inflation in a closed LQC universe that is sourced by a single scalar field. We focus on initial conditions which in the classical theory do not lead to an inflationary spacetime. The initial conditions are imposed in the Planck regime considering a small homogeneous patch of the universe. For such a patch the spatial curvature term plays an important role and inflation can only take place when the initial conditions are selected such that the potential energy of the scalar field is dominant at the initial time and is large enough to overcome the spatial curvature of the closed universe. When the initial energy density is dominated by the kinetic energy of the inflaton field, the universe recollapses before  inflation can occur due to the curvature of a closed universe. This results in a big crunch singularity following the recollapse.  One may be tempted to consider initial conditions corresponding to a very large initial volume such that the effect of the spatial curvature becomes so small that the pre-inflationary branch has no recollapse. However, this requires assuming an unnaturally large initial homogeneous patch of the universe in order to set initial conditions for inflation in the Planck regime. Typical initial conditions in the Planck regime start with patches which are not assumed to be homogeneous at macroscopic scales and these are the ones considered in our analysis.

In the following, we present two representative sets of initial conditions in the parameter space. The first set of initial conditions has the potential energy dominant at the bounce but does not allow for inflation in the classical theory because the potential energy is unable to overcome the classical recollapse caused by the spatial curvature term. The second representative set of initial conditions corresponds to initially  dominant kinetic energy which again results in a classical recollapse and a big crunch singularity. For both sets we find that because of quantum gravitational effects, big crunch singularities are avoided and inflation occurs after a few cycles of expansion and contraction. 

 In our numerical analysis, the fundamental equations of motion are the effective Hamilton's equations  (\ref{vdot})-(\ref{phidot}) introduced in Sec. \ref{review}. The initial conditions are chosen such that the universe has the highest allowed density given its volume at $t=0$.  In general, the initial conditions that must be specified are the values of the phase space variables $v$, $b$, $\phi$ and $p_\phi$ at $t=0$. (The initial conditions will be labelled by the subscript `0'.) In our simulations, we choose the initial volume $v_0$ and the initial value of the scalar field $\phi_0$ as the two initial free parameters. Since the initial conditions are chosen at a point where the energy density is maximal, the conjugate momentum of the scalar field can be determined by 
\bq
p_{\phi,0}=\pm v_0\sqrt{2\rho_\mathrm{max}-2U(\phi_0)},
\eq
where the `$\pm$' reflects the two possible signs of the initial velocity of the scalar field. Finally, the momentum $b_0$ is fixed by the vanishing of the Hamiltonian constraint (\ref{hamiltonian}), given $v_0$, $\phi_0$ and $p_{\phi,0}$. Of the possible solutions for $b_0$, we used the positive solution 
for all the plots shown below. Finally, all our simulations were performed using a combination of the StiffnessSwitching, ExplicitRungeKutta, and Automatic numerical integration solving methods in Mathematica with the precision and accuracy goals set to 11.

\subsection{Representative initial conditions}

\begin{figure}
\includegraphics[width=8cm]{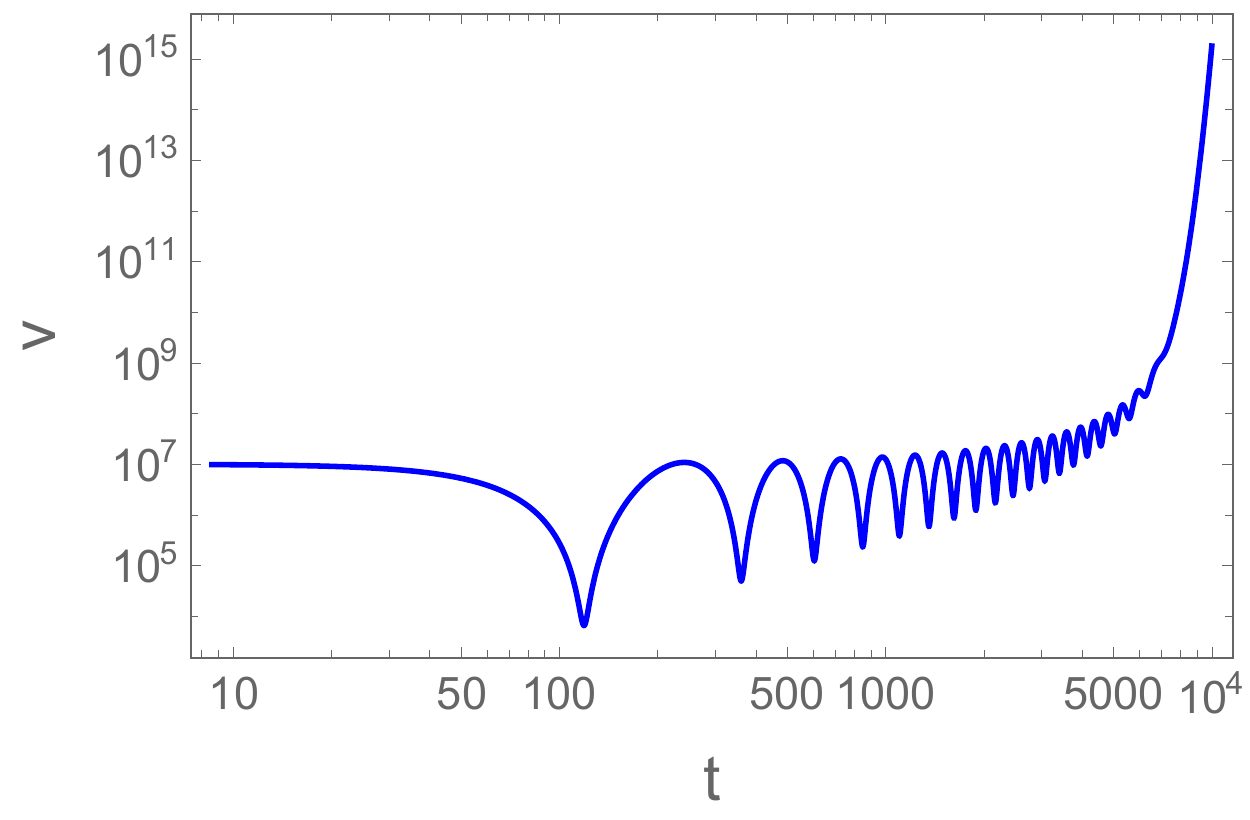}
\includegraphics[width=8cm]{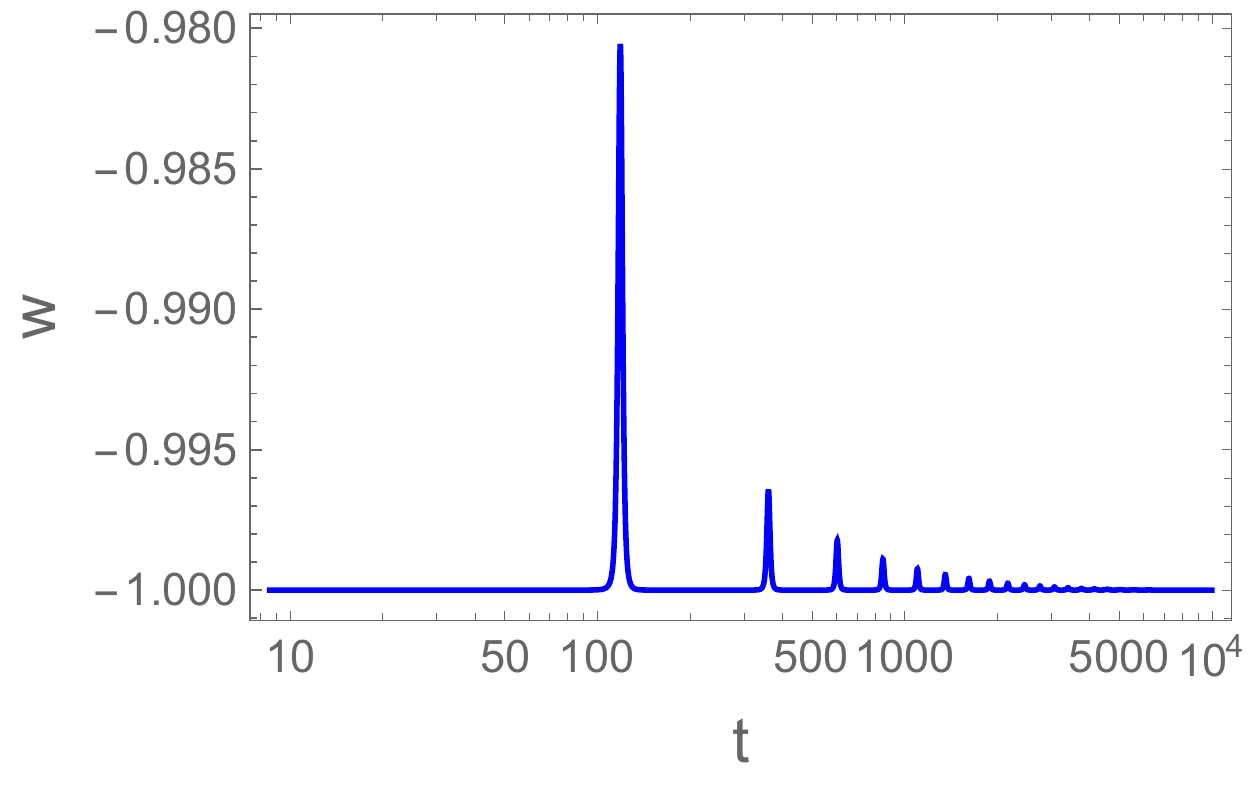}
\includegraphics[width=8cm]{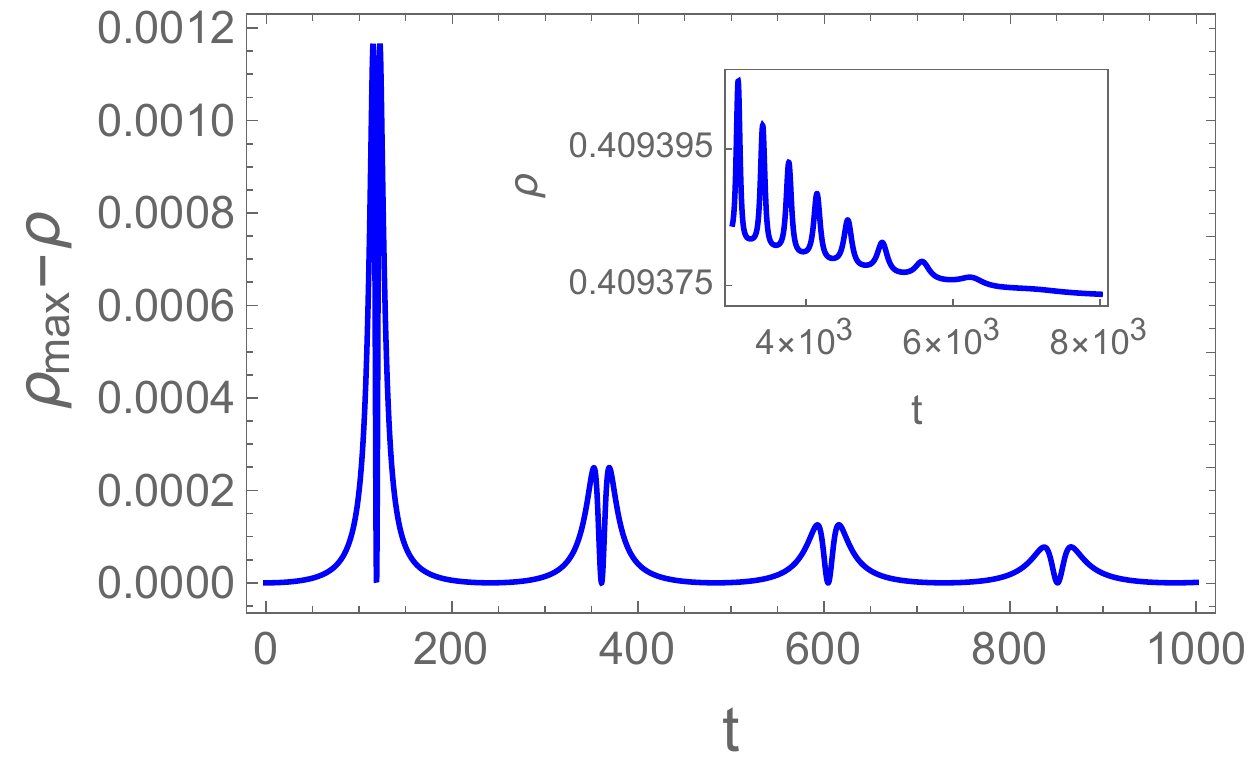}
\includegraphics[width=8cm]{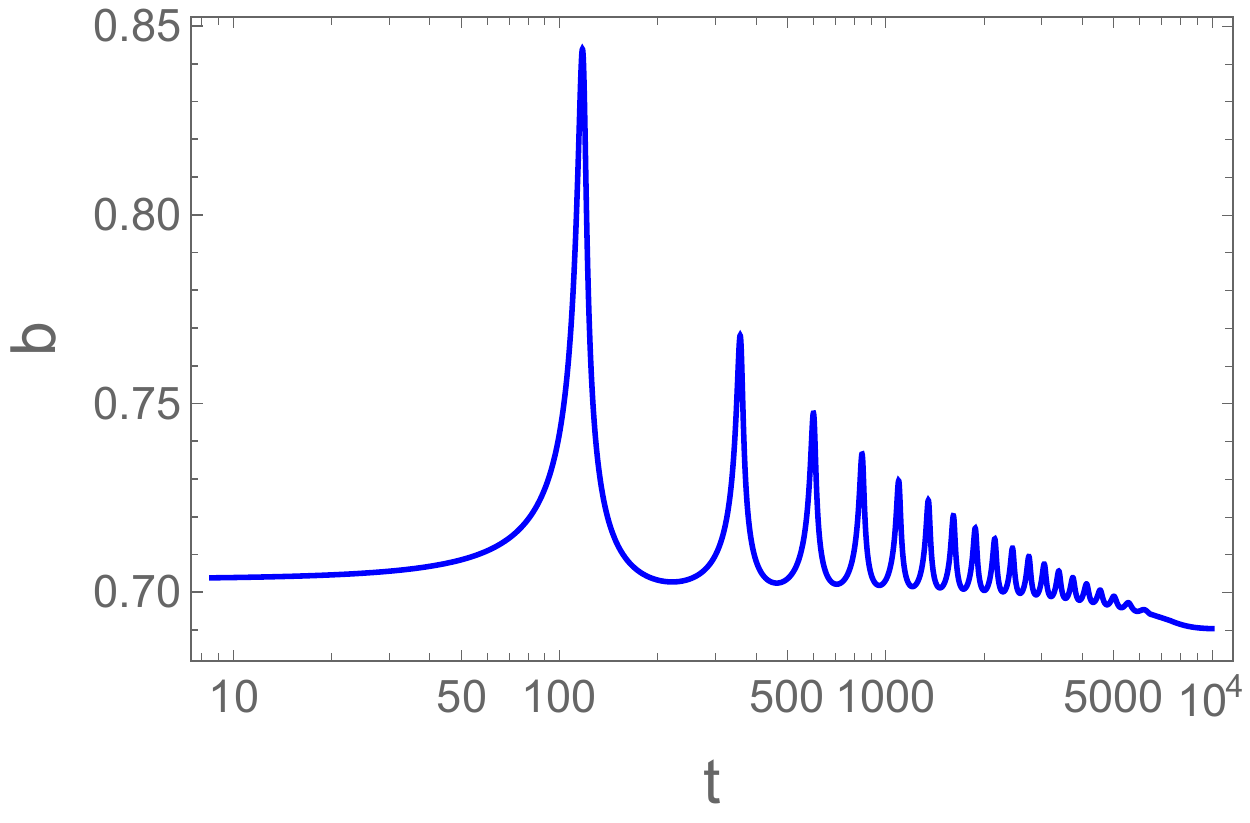}
\caption{In this figure, the time evolution of the configuration variables, the equation of state and the energy densities are depicted for the initial conditions (\ref{initial1}). The initial energy density is chosen to be the maximum allowed density and is completely dominated by the potential energy. All the variables except the scalar field exhibit oscillatory behavior as the universe undergoes a series of bounces and recollapses. The energy density reaches its maximum value at each bounce and recollapse. Inflation happens in the Planck regime when the energy density is about $0.41$ as the inflaton slowly rolls down the right wing of the chaotic potential.}
\label{f1}
\end{figure}

Below we discuss two representative cases for the inflationary potential $U = \tfrac{1}{2} m^2 \phi^2$, one with the potential energy dominant at an initial time in the Planck regime and the other with the kinetic energy dominant. Since we are interested in cases in which inflation does eventually take place, the mass of the inflaton field for the chaotic potential will be fixed as in the standard inflationary paradigm using the scalar power spectrum $A_s$ and the scalar spectral index $n_s$ for the pivot mode \cite{Planck2018}
\bq
\lb{4.2}
\ln (10^{10}A_s)=3.044\pm0.014  ~(68\% \mathrm{CL}) ,\quad\quad n_s=0.9649\pm0.0042 ~(68\% \mathrm{CL}) .
\eq
Using these the mass of the scalar field is fixed to $m=1.23\times 10^{-6}$ in this subsection.

The first set of representative initial conditions we would like to discuss is presented in Fig. \ref{f1}. This corresponds to
\bq
\lb{initial1}
v_0=10^7, \quad \quad \phi_0=7.33\times10^5,
\eq
in Planck units. With these initial conditions, at initial time $t_0=0$, the initial velocity of the scalar field is zero. Though the inflaton starts with all its energy in potential energy, such a universe soon recollapses and encounters a big crunch singularity in GR.  In contrast, we find that in LQC the universe undergoes a series of bounces and recollapses. These cycles occur before the conditions become favorable for inflation to begin. We find that both the maxima and minima of the volume of the universe increase with each bounce following the initial recollapse at $t=0$. Note that this case provides an example of the type of universe discussed in Sec. II where the maximal energy density corresponds to a recollapse rather than a bounce. In the current case, since the universe is initially dominated by potential energy, the equation of state is close to negative unity and thus at the maximum energy density,
\bq
\ddot v|_{\rho_\mathrm{max}} \approx - 12\pi G v \rho_2.
\eq
As a result, depending on the sign of $\rho_2$, the turning point at the maximum energy density can either be a bounce or a recollapse.  From the $\rho_\text{max}-\rho$ and $b$ plots of Fig. \ref{f1}, we see that the bounces and recollapses occur at the maximum energy density allowed at that time. It is evident by comparing the $\rho_\text{max}-\rho$ plot to the volume plot above it that this difference vanishes whenever the universe reaches a turnaround point, corresponding to the maximum density being achieved. Moreover, since $\ddot v$ and $\dot b$ have the opposite signs at $\rho=\rho_\mathrm{max}$, the sign of $\ddot v$ changes between each bounce and recollapse, which is consistent with the  oscillating behavior of the momentum $b$. The peaks of $b$ also show a noticeable decreasing hysteresis-like behavior \cite{ds2020}. This behavior is also seen in the increase in the maxima (also the minima) of the volume in subsequent cycles. 

When the universe is in the cyclic phase, the equation of state $w = P/\rho$ 
also changes periodically. It is interesting to note that although the universe is initially dominated by the potential energy of the scalar field, corresponding to $w=-1$, inflation does not take place immediately after the first bounce but rather at around $t=10^4$ (in Planck seconds). The cyclic phase of the universe is accompanied by oscillatory behavior in the equation of state, which attains a local maximum at each bounce and a local minimum at each recollapse. Furthermore, as seen from the behavior of  the energy density, we find that  inflation starts when the energy density is almost Planckian. Thus, for the above initial conditions we find a successful onset of inflation because of loop quantum gravitational effects, even though this universe is unable to inflate in the classical theory due to the big crunch singularity.

\begin{figure}
\includegraphics[width=8cm]{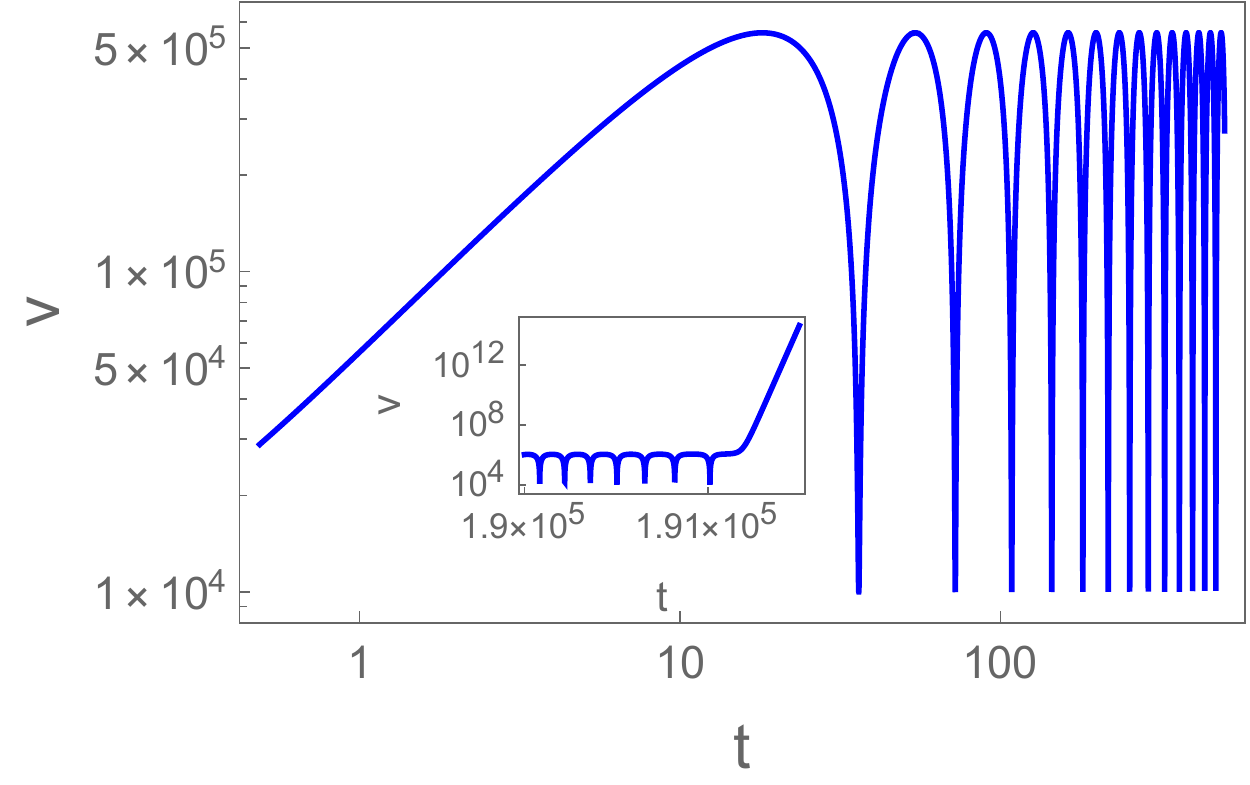}
\includegraphics[width=8cm]{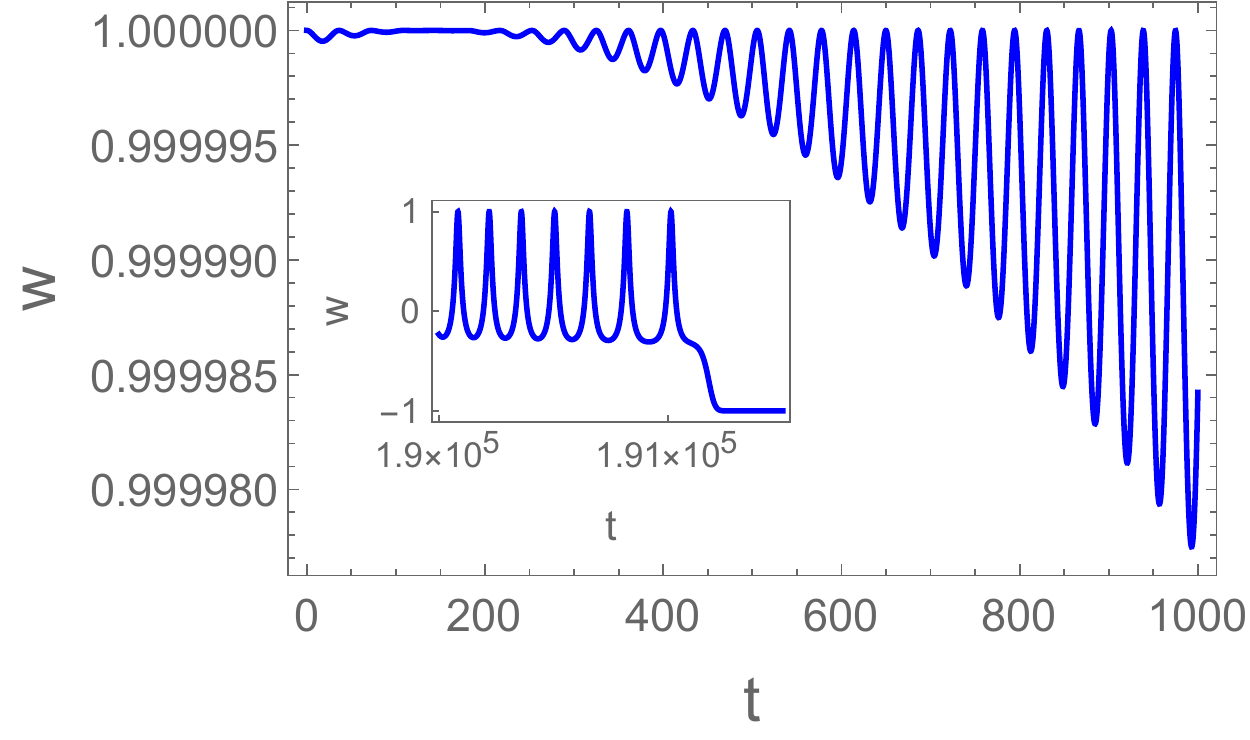}
\includegraphics[width=8cm]{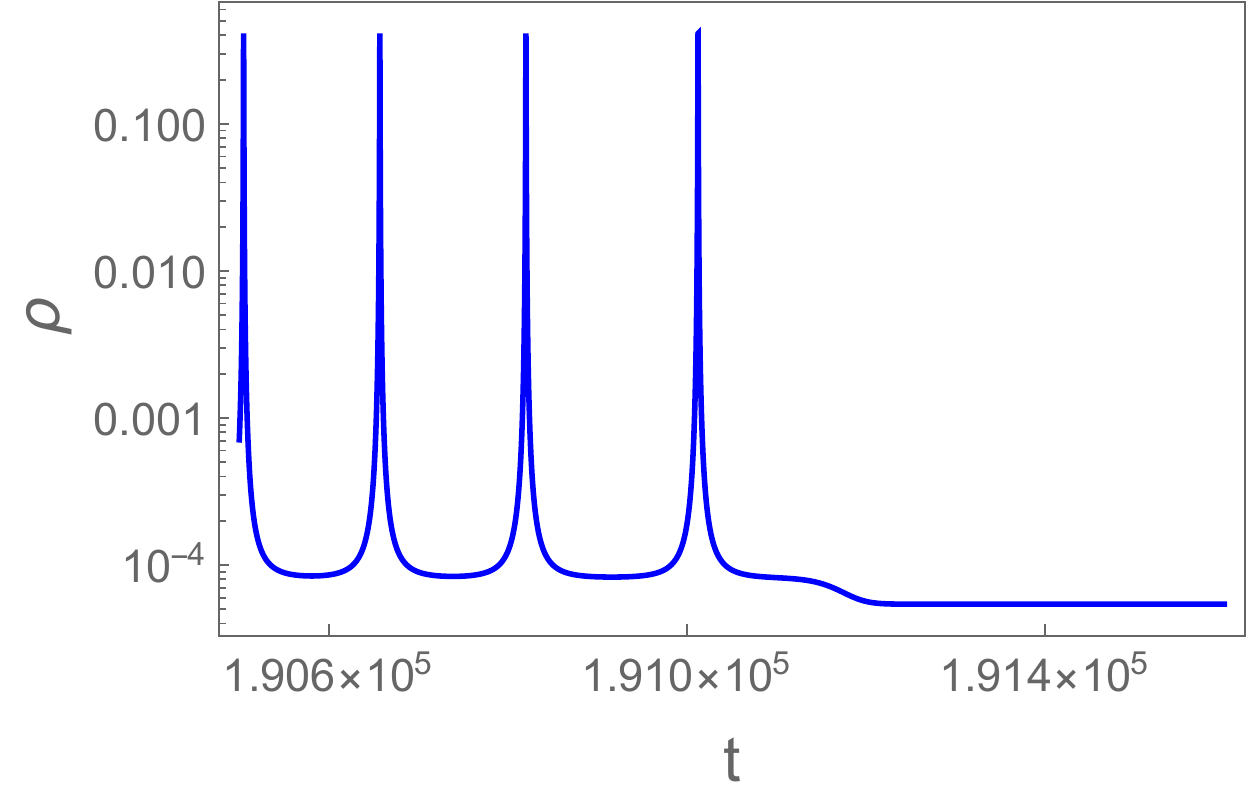}
\includegraphics[width=8cm]{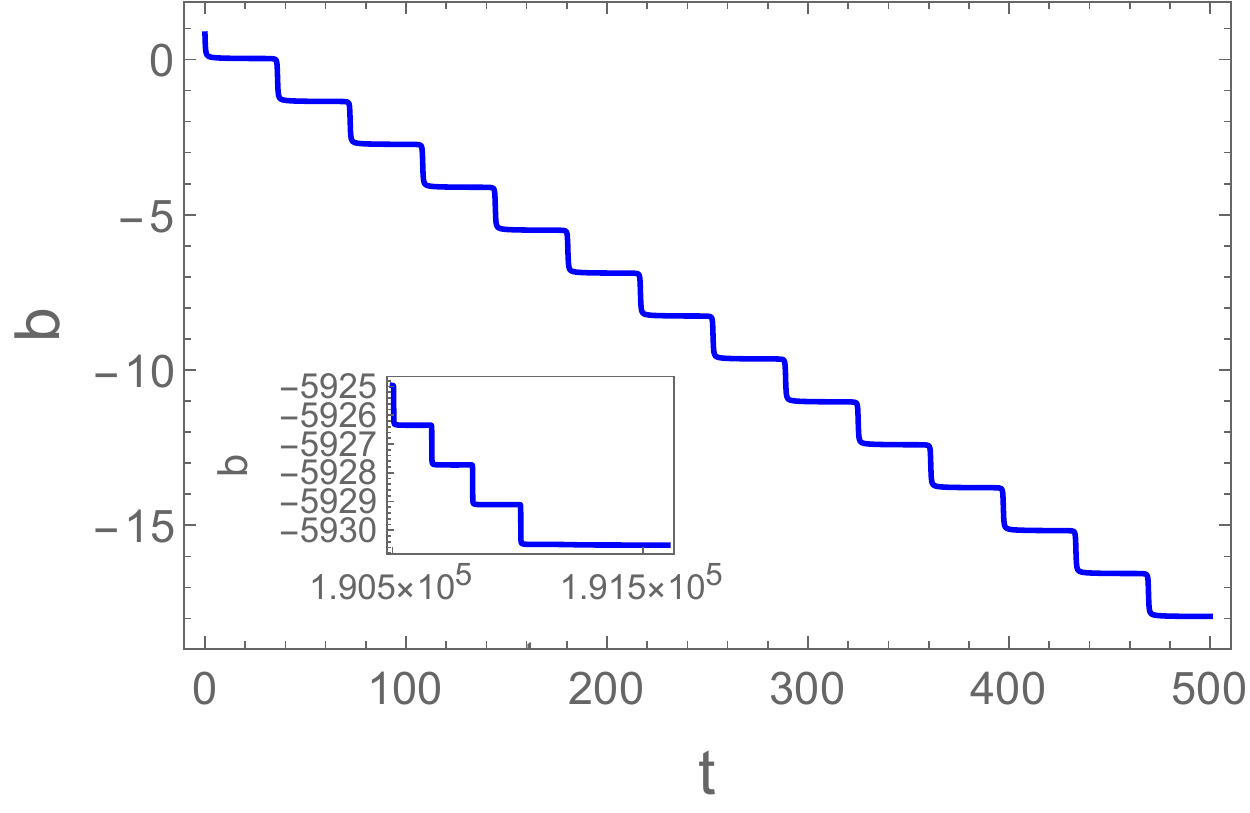}
\caption{With the initial conditions given in (\ref{initial2}) and a kinetic energy dominated initial state, the volume, the equation of state, the energy density and the momentum $b$ are shown from the initial bounce through the first few cycles. The inset plots in these subfigures show the behavior of the respective variables near the onset of inflation at around $1.91\times10^5$ Planck seconds. For the plot of the energy density, its evolution at late times is displayed. In this figure, the bounces occur at the maximum energy density while the recollapses occur at the minimum energy density. }
\label{f2}
\end{figure}

The second example is depicted in Fig. \ref{f2} with the initial conditions given by 
\bq
\lb{initial2}
v_0=10^4, \quad \quad \phi_0=7.36,
\eq
in Planck units. These correspond to a negative initial velocity for the scalar field. As in the previous case, the positive curvature slows the expansion of the universe after $t=0$, which in this case is a bounce,  leading to a recollapse which is followed by a non-singular  bounce in LQC. Initially the universe undergoes cycles of expansion and contraction with little change in the maximum volume. It is at time  $t=300$ that the recollapse volume begins to significantly increase with each cycle. During this phase, with each subsequent recollapse, the potential energy fraction gets higher, corresponding to the equation of state reaching a lower minimum value. The time it takes to complete a cycle also tends to increase with time, with the next recollapse happening a little later after the bounce than the one before it. A noticeable hysteresis-like phenomenon can be observed in the $w$ plot where all the troughs correspond to recollapse points whose values decrease with time, while all the peaks correspond to bounce points, which remain at unity throughout the pre-inflationary evolution. We see that the minimum of the equation of state decreases during each cycle until $w$ reaches $-1/3$ after one final bounce and then inflation takes place, preventing any further recollapses.

In the figure, we only display  the volume, equation of state, and momentum $b$ corresponding to the first several cycles. The behavior of these variables at the onset of inflation is depicted in the inset plots. Note that the momentum $b$ behaves differently than in the first case. In this case, $b$ is monotonically decreasing. As a result, $\ddot v$ is always negative when $\rho=\rho_\mathrm{min}$ and positive when  $\rho=\rho_\mathrm{max}$. That is, in contrast to Fig. \ref{f1}, each bounce happens at the maximum energy density while each recollapse happens at the minimum energy density. 
\begin{figure}[tbh!]
\includegraphics[width=220pt]{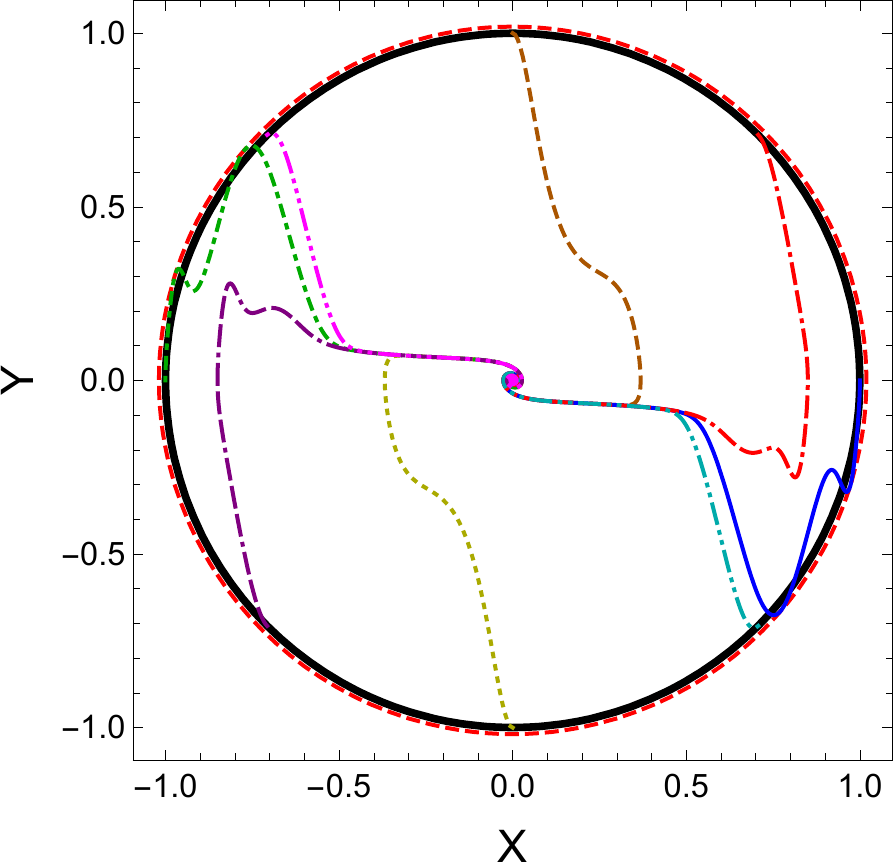}
\includegraphics[width=220pt]{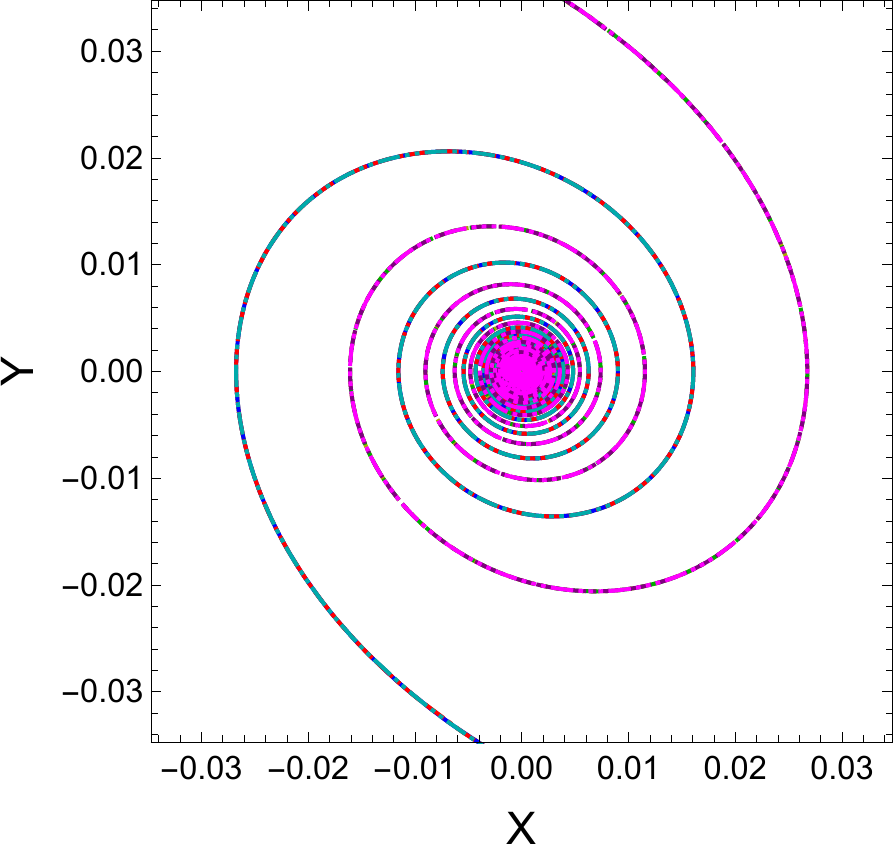}
\includegraphics[width=220pt]{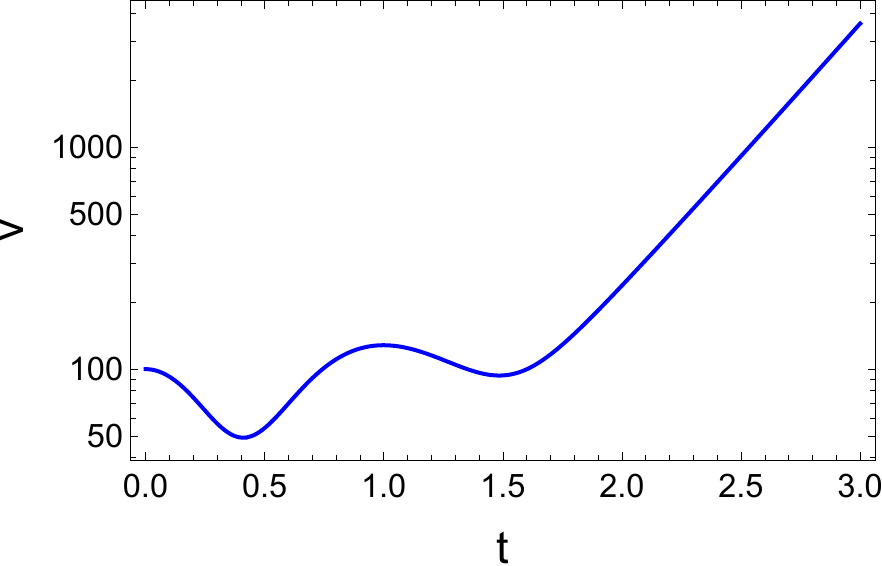}
\includegraphics[width=220pt]{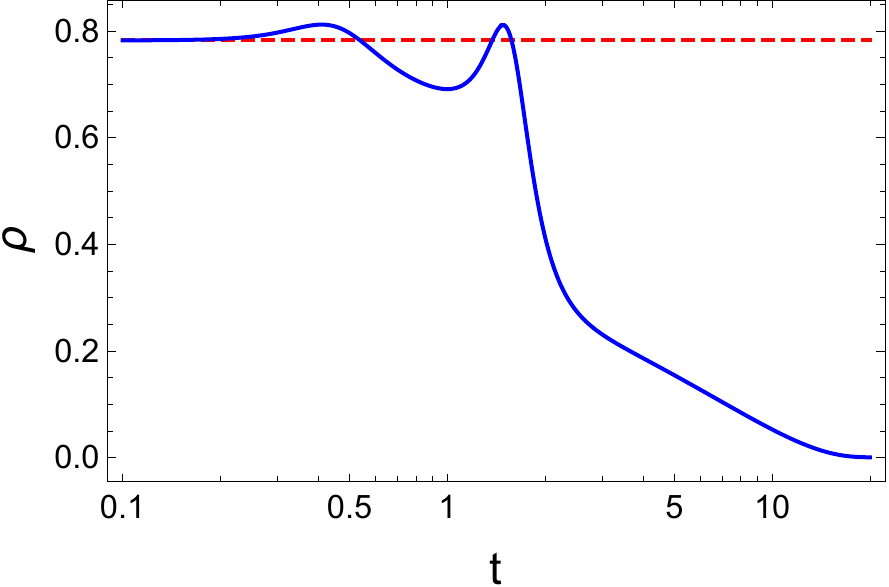}
\caption{Phase space portrait along with volume and density plots for the chaotic potential with $m=0.50$ and initial volume $v_0=100.00$. The upper left plot shows the entire phase space region. The solid black curve corresponds to the energy density at $t=0$ where the initial conditions are set, and the dashed red curve represents the maximum energy density that is achieved during the time evolution of any of these initial conditions, which just slightly exceeds the initial energy density. The upper right plot zooms in on the spiral structure. The two lower plots display the volume and energy density, respectively, for the initial conditions $(X_0=1,Y_0=0)$, corresponding to the blue solid curve in the phase space plot. The dashed red straight line in the bottom right panel represents the energy density at which the initial bounce takes place. }
\label{f3}
\end{figure}
\subsection{Phase space portraits}
\begin{figure}
\includegraphics[width=220pt]{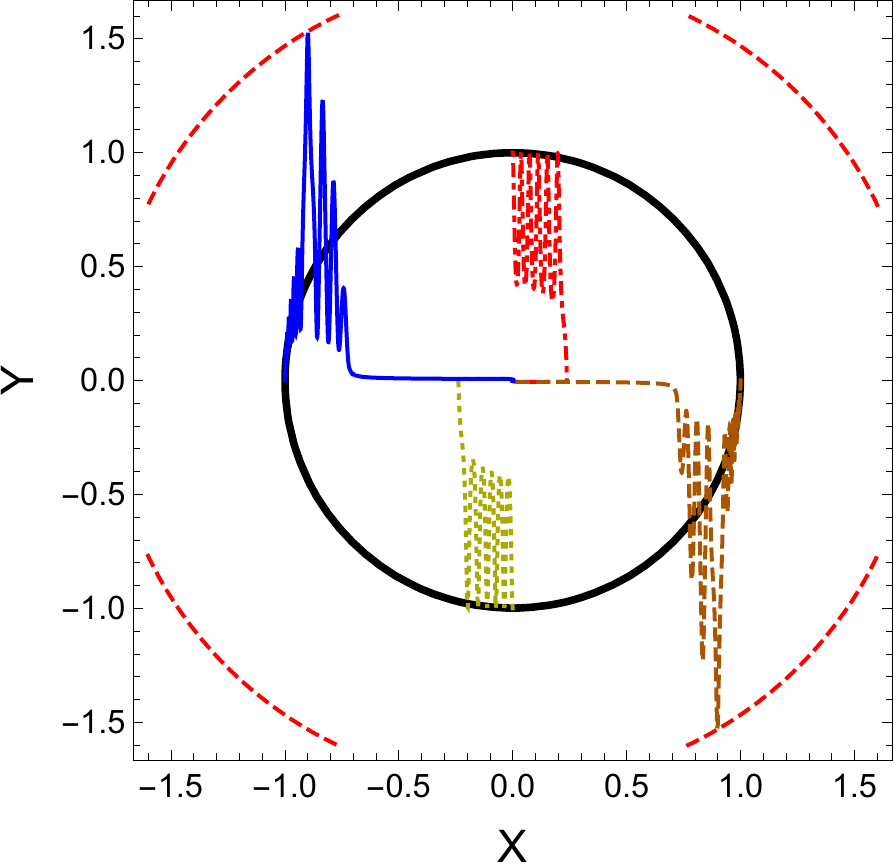}
\includegraphics[width=220pt]{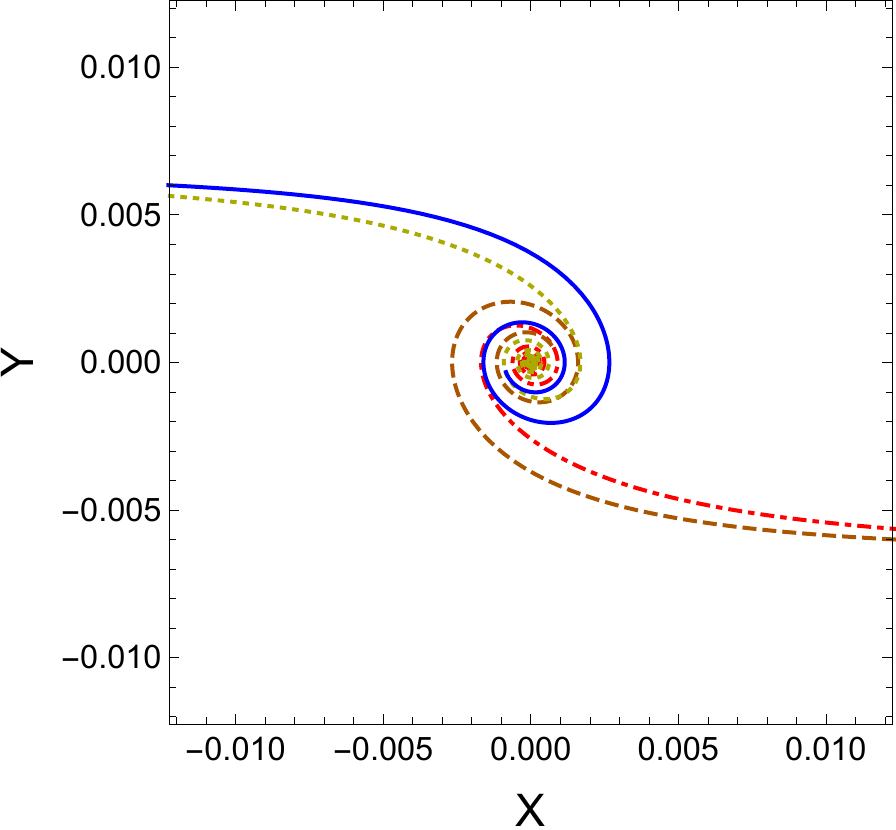}
\includegraphics[width=220pt]{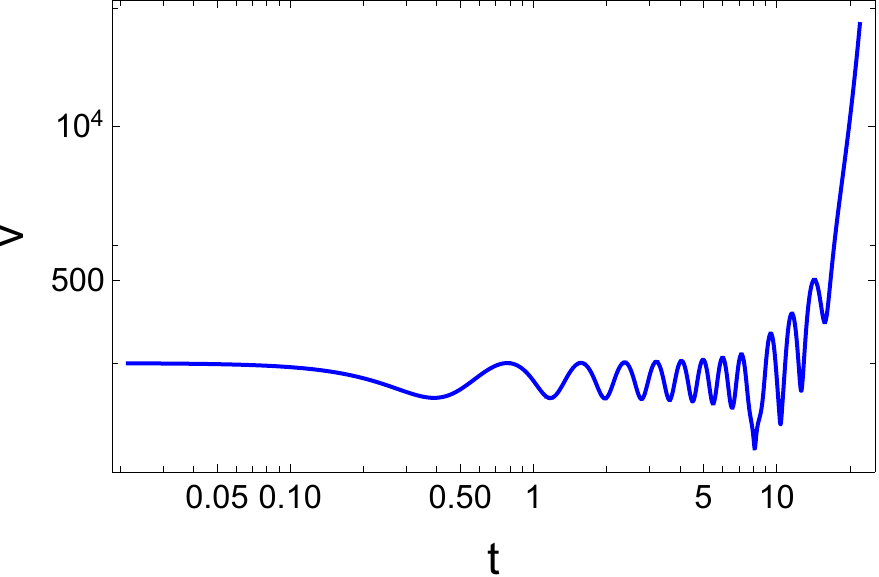}
\includegraphics[width=220pt]{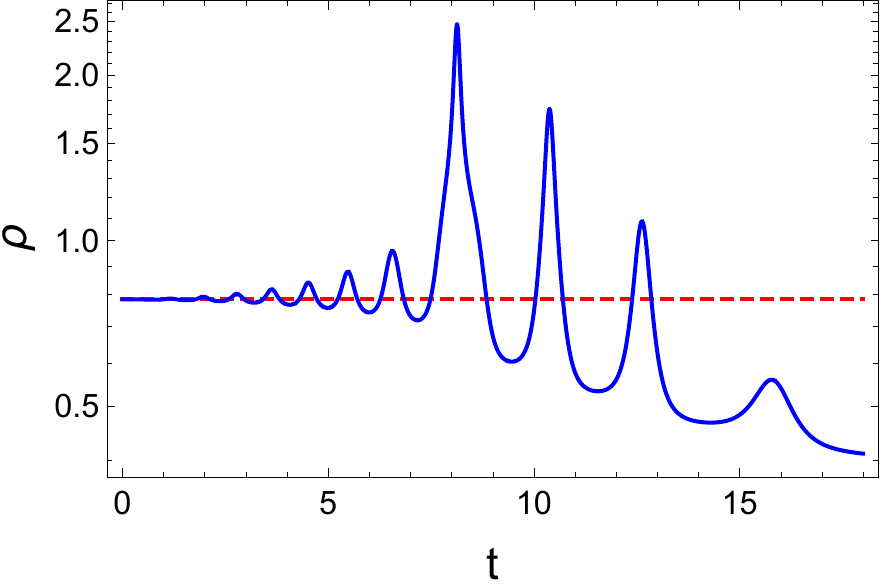}
\caption{Phase space portrait along with volume and density plots for the chaotic potential with $m=0.05$ and initial volume $v_0=100$. The upper left plot shows the entire phase space region with the dashed red circle showing the maximum energy density that is achieved during the evolution of two of the solutions, which is significantly larger than the initial energy density represented by the thick black circle. The upper right plot zooms in on the reheating phase. The two lower plots display the volume and energy density of the blue solid curve in the phase space plot. In the volume plot on the lower left there are twelve bounces before the inflationary phase begins. In the density plot on the lower right the energy density during many of the bounces exceeds the initial energy density, represented by the red dashed line. These bounces correspond to peaks outside of the black circle in the top left phase space plot. The universes corresponding to the red (dot-dashed) and light green (dotted) curves also undergo several bounces, but during these bounces the energy density is near the maximal value, unlike for the blue and brown curves.} 
\label{f4}
\end{figure}

In this subsection, we present phase space portraits which are used to understand some aspects of the qualitative behavior of the numerical solutions for a variety of distinct initial conditions. These phase space portraits are based on a set of first-order ordinary differential equations that are equivalent to the Hamilton's equations for the scalar field in (\ref{phidot}). More specifically, in order to make sure that all the initial conditions are selected from the unit circle representing the bounce point at the initial time, the following phase space variables are used in the phase space portraits, namely, 
\bq
\lb{3b1}
X=\frac{m \phi}{\sqrt{2\rho_{\mathrm{max},0}}}, \quad \quad Y=\frac{\dot \phi}{\sqrt{2\rho_{\mathrm{max},0}}} ~.
\eq
At the initial time these satisfy the condition 
\bq
\lb{3b2}
X^2_0+Y^2_0=1,
\eq
and obey the equations of motion
\bqn
\lb{3b3}
\dot X &=&m Y ,\\
\lb{3b4}
\dot Y&=&- m X-3 HY . 
\eqn
It should be noted that the variables $X$ and $Y$ and their respective dynamical equations (\ref{3b3})-(\ref{3b4}) do not form a closed system as the Hubble rate $H$ in a closed universe cannot be expressed solely as a function of $X$ and $Y$ because of the presence of spatial curvature.  As a result,  the Friedmann equation  (\ref{friedmann}) should be added to form a closed system described by the variables $X$, $Y$ and $v$. Since we are interested in understanding inflationary attractors, we focus on the phase space portraits in the subspace spanned by $X$ and $Y$. Furthermore, since the purpose of this section is to investigate the qualitative behavior of the solutions, in order to achieve a faster convergence of the solutions to the attractors in the plots, a fictitious mass that is much larger than the actual mass (determined from the inflationary scenario and CMB data) is used. For chaotic inflation, we show two representative phase space portraits which are Figs. \ref{f3}-\ref{f4}. In the former figure, the universe undergoes a few bounces before inflation takes place, while in the latter, there is a greater number of bounces and recollapses before inflation occurs.

Fig. \ref{f3} displays the time evolution of eight distinct solutions, which are shown in different colors and styles from the initial time to the reheating phase.  Since the initial maximum energy density is determined by the initial volume (fixed to $v_0=100$), the solutions in the figure start from the same density and hence the same solid black circle but differ in the initial value of the scalar field and its time derivative. There are trajectories originating from a potential energy dominated bounce/recollapse, such as the green dot-dashed and blue solid curves, and trajectories originating from a kinetic-dominated bounce, such as the brown dashed and yellow dotted curves. Other trajectories such as the dot-dash-dashed red and dot-dash-dash-dashed purple curves start from a bounce where the potential energy and the kinetic energy are comparable in magnitude. All these trajectories have qualitatively similar behavior in the sense that after a few bounces and recollapses, the curves starting from each half of the circle merge into horizontal lines (these are almost parallel to the x axis and are known as inflationary separatrices). Subsequently, the two horizontal lines merge in a spiral at the center of the portrait. The spiral structure corresponds to the reheating phase where all the trajectories overlap with one another as shown in the right panel of Fig. \ref{f3}. In this regime, the scalar field behaves like a damped harmonic oscillator, and solutions starting from different initial conditions in the Planck regime result in the same classical evolution. The origin $(X_0=0, Y_0=0)$ is a fixed point of the system as it is a static solution of the dynamical equations (\ref{3b3})-(\ref{3b4}).

In the bottom panels, we have explicitly shown the time evolution of the volume and energy density for the set of initial conditions with $(X_0=1, Y_0=0)$. This describes a universe that starts from a potential energy dominated state and after two bounces quickly enters into a phase of inflation when the energy density of the inflaton field is still in the Planck regime. We find that initially, the universe is in a state of contraction. When the volume of the universe reaches about 50 (in Planck volume), a quantum bounce takes place and the universe enters an expanding phase. The bottom left panel of Fig. \ref{f3} explicitly shows the first and second bounces after the initial recollapse. The volume at the second bounce is around $90$. Since the volumes at the first and second bounces are smaller than the initial volume, the energy density at these bounces exceeds the initial energy density as is depicted in the bottom right panel of the figure. 

The next example we want to analyze is Fig. \ref{f4} where four distinct solutions are shown. Two of the solutions with $(X_0=\pm1, Y_0=0)$ start from  potential energy domination, while the other two with $(X_0=0,Y_0=\pm1)$ correspond to kinetic energy domination. All four solutions undergo a number of bounces and recollapses before inflation takes place. The qualitative dynamics of the solutions after the onset of inflation in Fig. \ref{f4} is the same as for those in Fig. \ref{f3}. In Fig. \ref{f4}, we also observe inflationary separatrices in the top left panel as well as a spiral near the origin in the top right panel where four distinct trajectories converge, showing that all the solutions result in a  classical reheating phase. Distinctive features in the pre-inflationary regime can be observed in the bottom panels where we focus on the time evolution of the blue  solid curve after the initial recollapse. In the bottom left panel, we see that there are altogether 12 bounces after the initial recollapse.  As compared with Fig. \ref{f3}, the change in the maximum energy density from one bounce to the next is much larger in Fig. \ref{f4}, which makes the red dashed circle in the top left panel be separated from the black circle. It can also be observed from the phase space portraits that the inflationary period is much longer for the potential energy dominated initial conditions than for the kinetic energy dominated ones. The trajectories with the former initial conditions merge into inflationary separatrices earlier than those with the latter initial conditions. When the kinetic energy dominates initially, the bounces along the red dot-dashed and green dotted curves happen mostly near the black circle in the phase space portraits, indicating a weak hysteresis.

\section{Starobinsky inflation in a closed FLRW universe in LQC }
\label{starobinsky}
\renewcommand{\theequation}{4.\arabic{equation}}\setcounter{equation}{0}
In the previous section we considered the case of chaotic inflation and found that quantum gravity effects assist the onset of inflation in a closed universe. In this section we will study Starobinsky inflation in the same setting. Unlike in classical cosmology, in LQC Starobinsky inflation is not obtained from an $R^2$ term in the action,\footnote{The action in LQC yields higher order terms but in a Palatini framework \cite{olmo-ps}.} rather one generally takes as given the Starobinsky potential in effective dynamics \cite{bg2016,bg2016b,lsw2018b,lsw2019}, whose form is explicitly given by
\bq
U=\frac{3m^2}{32 \pi G}\left(1-e^{-\sqrt{\frac{16\pi G}{3}}\phi}\right)^2 .
\eq
The mass $m$ is fixed to $2.44\times10^{-6}$ from the scalar power spectrum and the spectral index given in (\ref{4.2}). To determine the value of the mass we assume that the pre-inflationary dynamics would not change the scalar power spectrum in a significant way. Note that unlike chaotic inflation, which can take place in the Planck regime, Starobinsky inflation can only occur on the right wing of the potential, which corresponds to an energy scale that is $10^{13}$ orders of magnitude lower than the Planck scale. As a result, in the classical theory, when starting from the Planck regime the universe inevitably recollapses before inflation sets in, resulting in a big crunch singularity. We now study how the dynamics change in LQC.

\begin{figure}
\includegraphics[width=220pt]{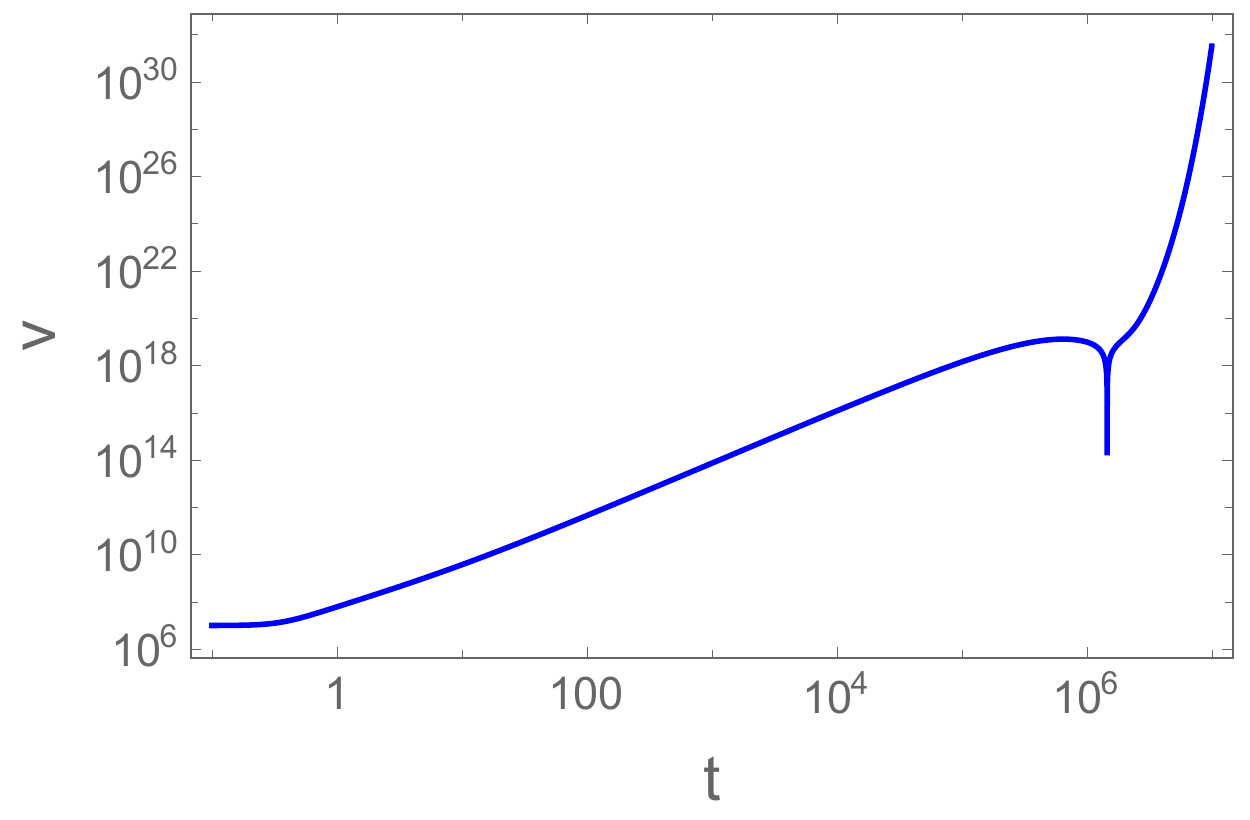}
\includegraphics[width=220pt]{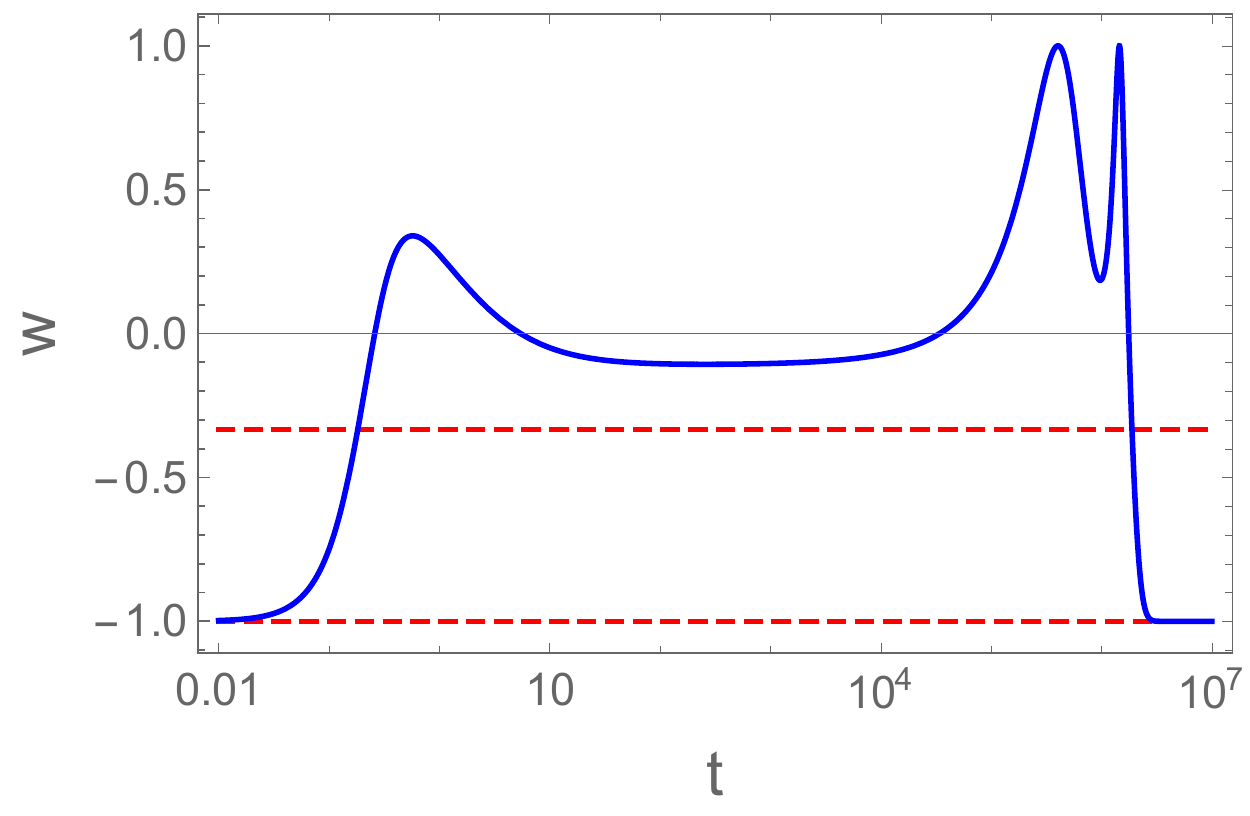}
\includegraphics[width=220pt]{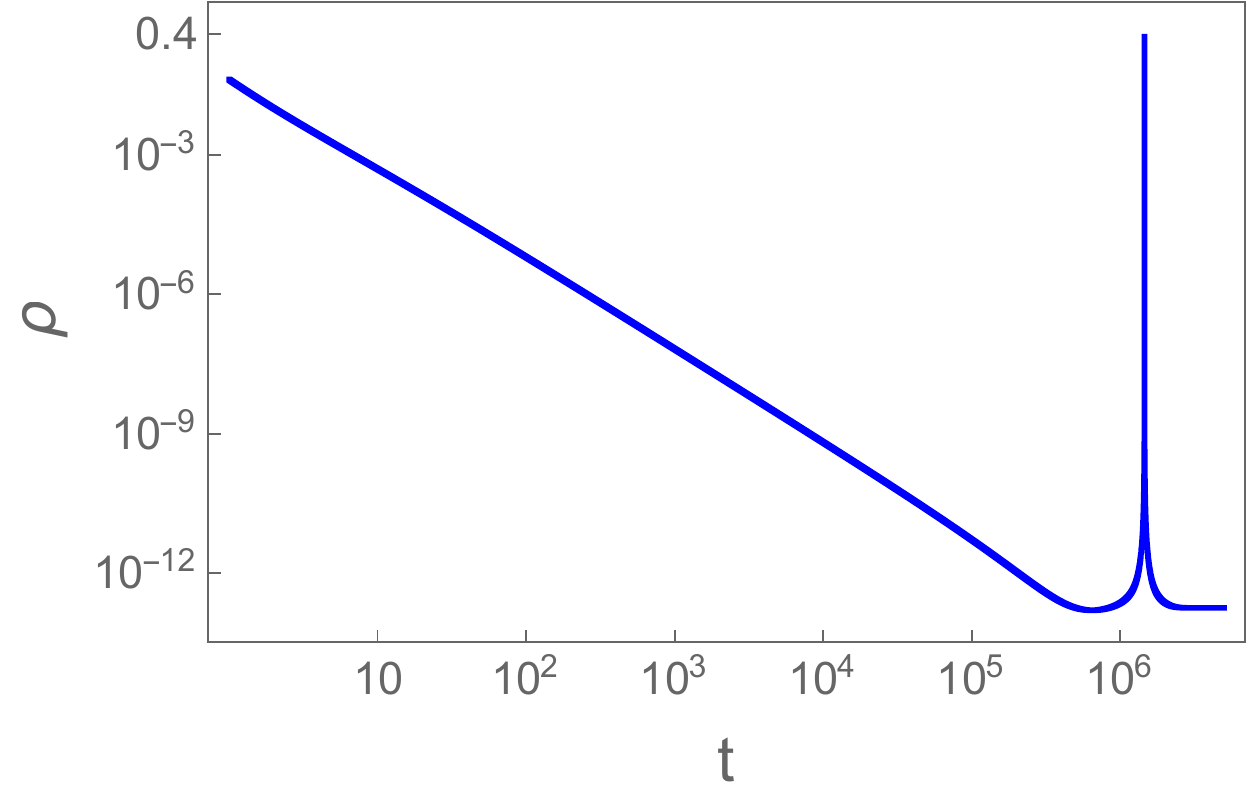}
\includegraphics[width=220pt]{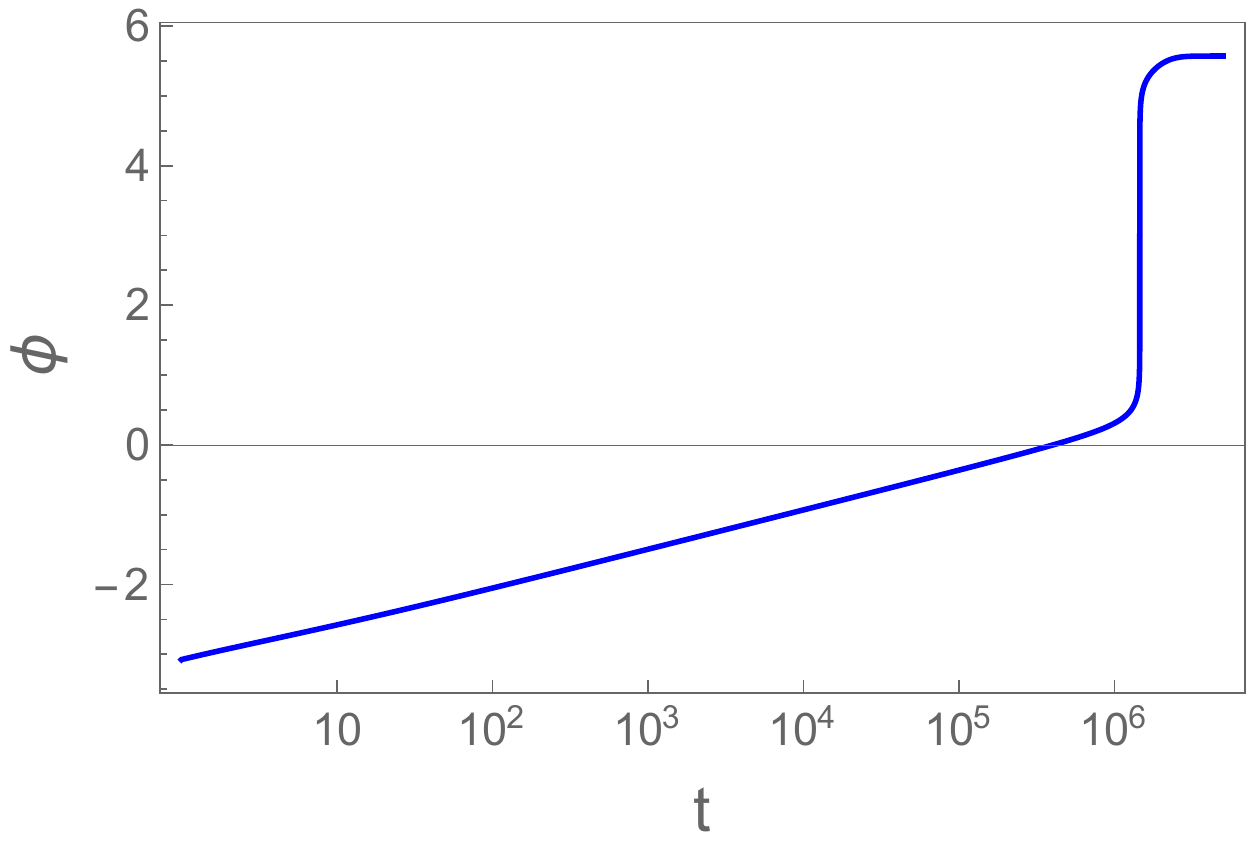}
\caption{This figure shows the volume, equation of state, density, and scalar field as a function of time for the Starobinsky potential with the initial conditions specified by (4.2). The initial conditions are set at the bounce, which takes place at the maximum allowed density. The initial energy density is all in the form of potential energy. This universe undergoes a single bounce before entering inflation. The inflaton rolls down the left wing of the potential, up the right wing, and then inflation occurs as it rolls down the right wing.}
\label{f5}
\end{figure}
\subsection{Some representative initial conditions}

The first example we would like to discuss corresponds to Fig. \ref{f5}, which results from the initial conditions (in Planck units)
\bq
\lb{initial3}
v_0=10^7,\quad \quad \phi_0=-3.48.
\eq
These initial conditions are chosen at a bounce with energy density $\rho_0=0.41$. In this case, the initial bounce is completely dominated by the potential energy of the scalar field and the inflaton is released from rest on the left wing of the potential, rolls down, and then climbs up the right wing until reaching the turnaround point. Inflation takes place at $\rho \approx 10^{-13}$ when the inflaton slowly rolls down the right wing of potential. Hence the scalar field behaves in the same way as in a spatially-flat universe. However, the behavior of other dynamical variables is quite different. From the volume plot in Fig. \ref{f5}, one can see that before the onset of inflation at around $t=8.39\times10^6$, the universe undergoes a bounce at $t=1.46\times10^6$ where the energy density reaches the maximum allowed value and a recollapse at around $t=6.47\times10^5$ with the minimum allowed energy density at that time.

\begin{figure}
\includegraphics[width=220pt]{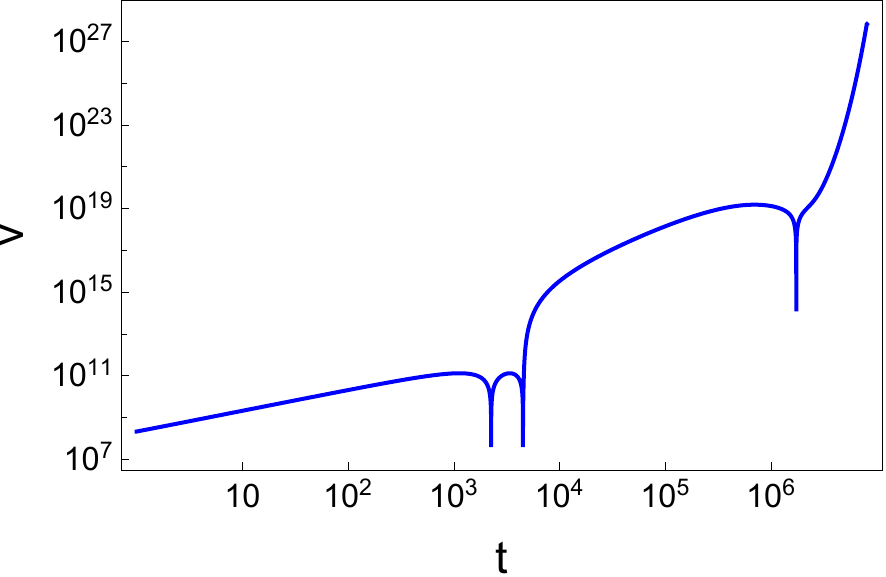}
\includegraphics[width=220pt]{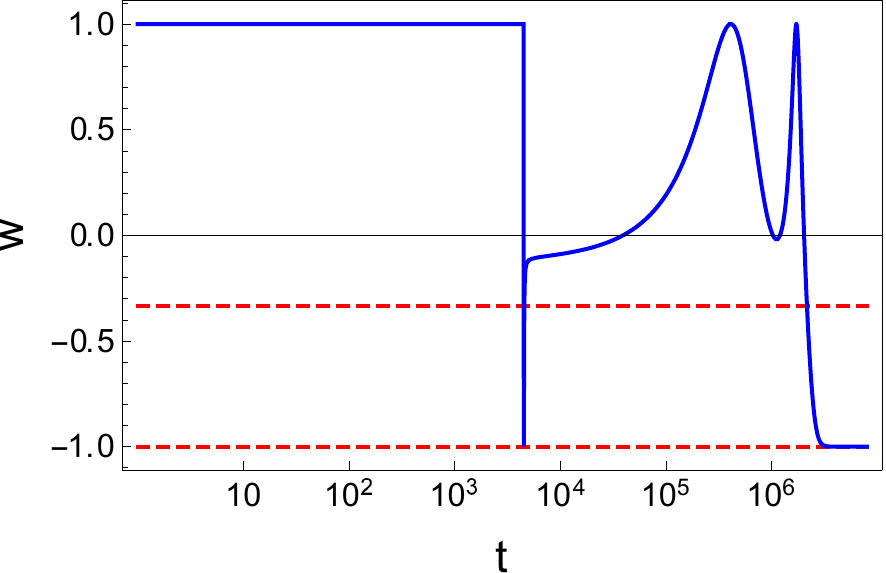}
\includegraphics[width=220pt]{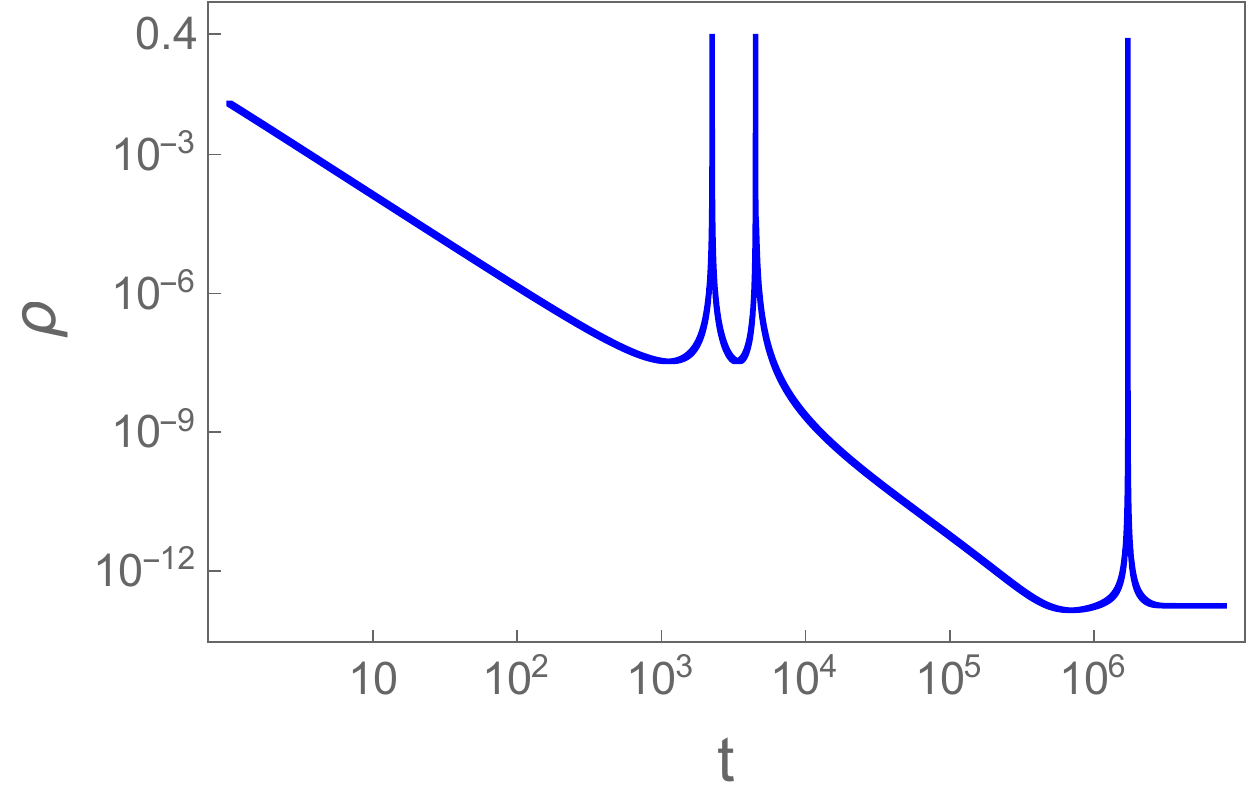}
\includegraphics[width=220pt]{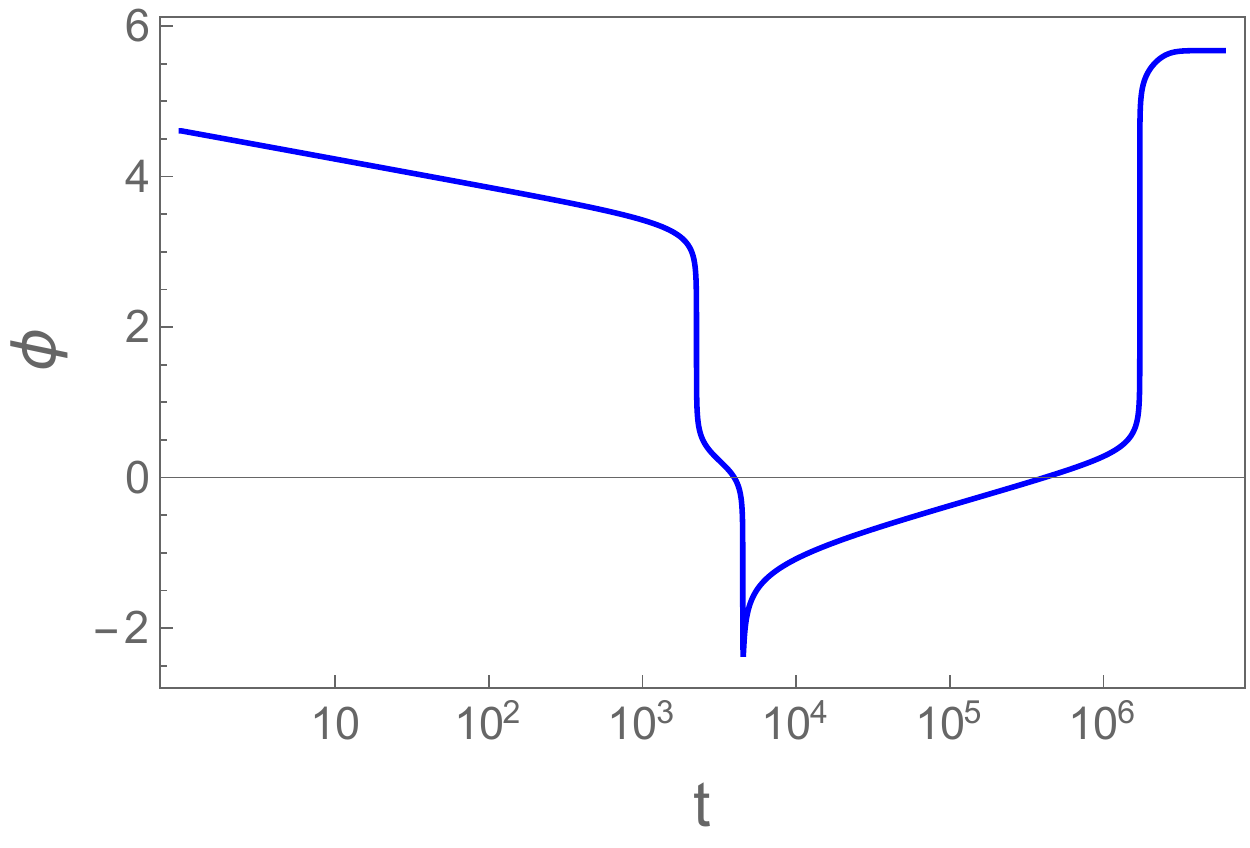}
\caption{This figure shows the volume, equation of state, density, and scalar field as a function of time for the Starobinsky potential with the initial conditions specified by (4.3). The initial conditions are set at the bounce, which takes place at the maximum allowed density. The initial energy density is dominated by the kinetic energy. This universe undergoes three bounces before entering inflation. The inflaton rolls down the right wing of the potential, up and down the left wing, back up the right wing, and then inflation occurs as it rolls down the right wing.}
\label{f6}
\end{figure}

The equation of state in the upper right panel of Fig. \ref{f5} shows explicitly that the initial bounce is dominated by potential energy with $w \approx -1$. Meanwhile the single bounce that takes place during the evolution is dominated by the kinetic energy of the scalar field as it corresponds to the last peak in the $w$ plot before the onset of inflation. The second to last peak in the $w$ plot corresponds to the moment when the inflaton crosses the origin, where the potential energy vanishes, forcing $w=1$. It should be noted that even though the potential energy is initially dominant, inflation cannot occur on the left wing of the potential because it is too steep for the slow-roll to occur. Only the right wing of the Starobinsky potential can drive inflation in both closed and spatially-flat universes. This is also manifest in the phase space portrait shown later where only a single inflationary separatrix is observed. Since this universe only has a single bounce after the initial one, we do not see oscillatory behavior in $w$ before the onset of inflation.

The second example is given in Fig. \ref{f6}, corresponding to a  universe initially dominated by the kinetic energy of the scalar field. The initial conditions for the numerical simulation are set at the big bounce at $t=0$, given  explicitly  by 
\bq
\lb{initial4}
v_0=3.75\times10^7,\quad \quad \phi_0=5.00,
\eq
which implies that the inflaton is initially released from the right wing of the potential with a large initial velocity. From the bottom right panel of Fig. \ref{f6}, we see that the inflaton first rolls down the right wing of the potential,  then climbs up the left wing and momentarily stops at the turnaround point. Afterwards, it rolls down the left wing of the potential, climbs up the right wing and then reaches another turnaround point. Finally, slow-roll inflation takes place when the inflaton slowly rolls down the right wing of the potential for the second time. The energy density at which inflation occurs is about $10^{13}$ orders of magnitude below the Planck energy. Fig. \ref{f6} shows that as compared with the previous example, the current case is richer in terms of the pre-inflationary dynamics. Note that the three recollapses associated with the three bounces in this example all occur at an energy scale far below the Planck energy, while all the bounces happen at a similar energy density that is around $\rho_\mathrm{crit}$. From the figure one notes that the third cycle of expansion and contraction is highly asymmetrical, which is due to the asymmetry of the Starobinsky potential itself. For this particular cycle the expanding (contracting) phase takes place on the left (right) wing of the potential. From the $w$ plot in the top right panel, one can easily identify several important moments featuring the behavior of the scalar field before the onset of inflation. For example, the first trough at $w=-1$ before $t=10^4$ corresponds to the turnaround point of the inflaton on the left wing of the potential. While the first two bounces take place before this trough, the effective dynamics following the first trough are qualitatively similar to what happened in Fig. \ref{f5}. This is confirmed by the similar behavior of the $w$ plots in these two cases in two particular regions: around the second to last peak where the inflaton crosses the origin and around the final peak where the universe undergoes a bounce.

\begin{figure}
\includegraphics[width=220pt]{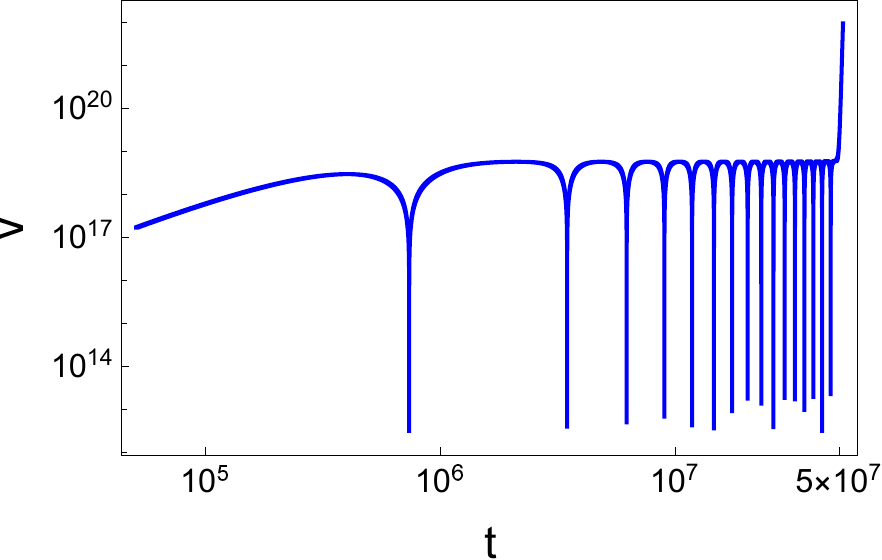}
\includegraphics[width=220pt]{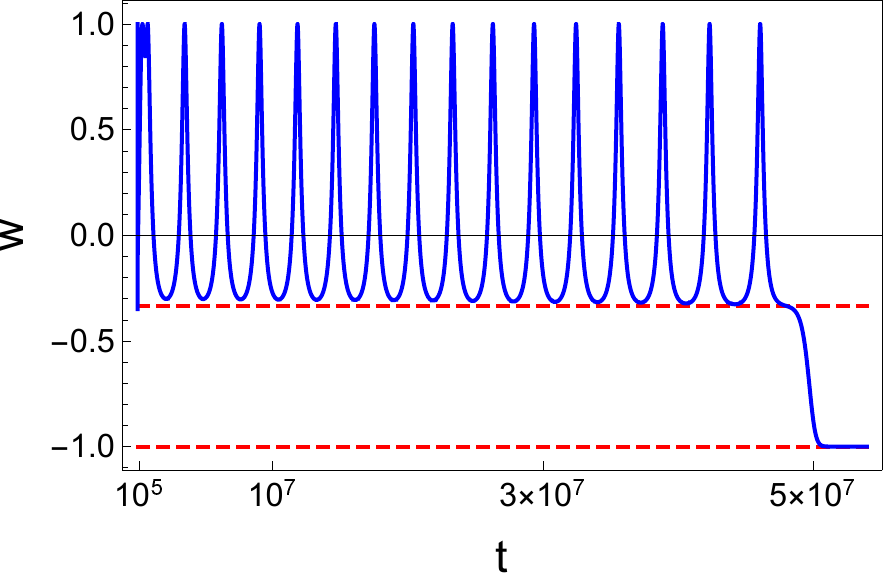}
\includegraphics[width=220pt]{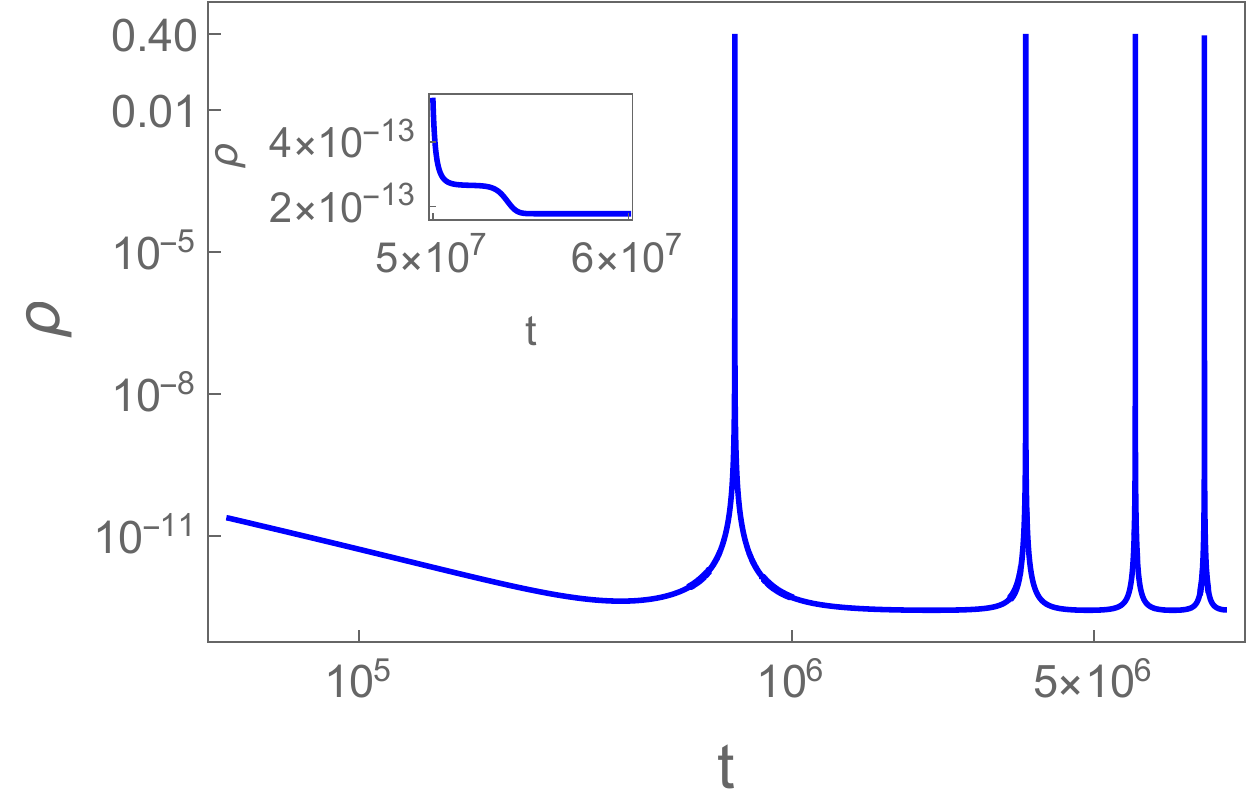}
\includegraphics[width=220pt]{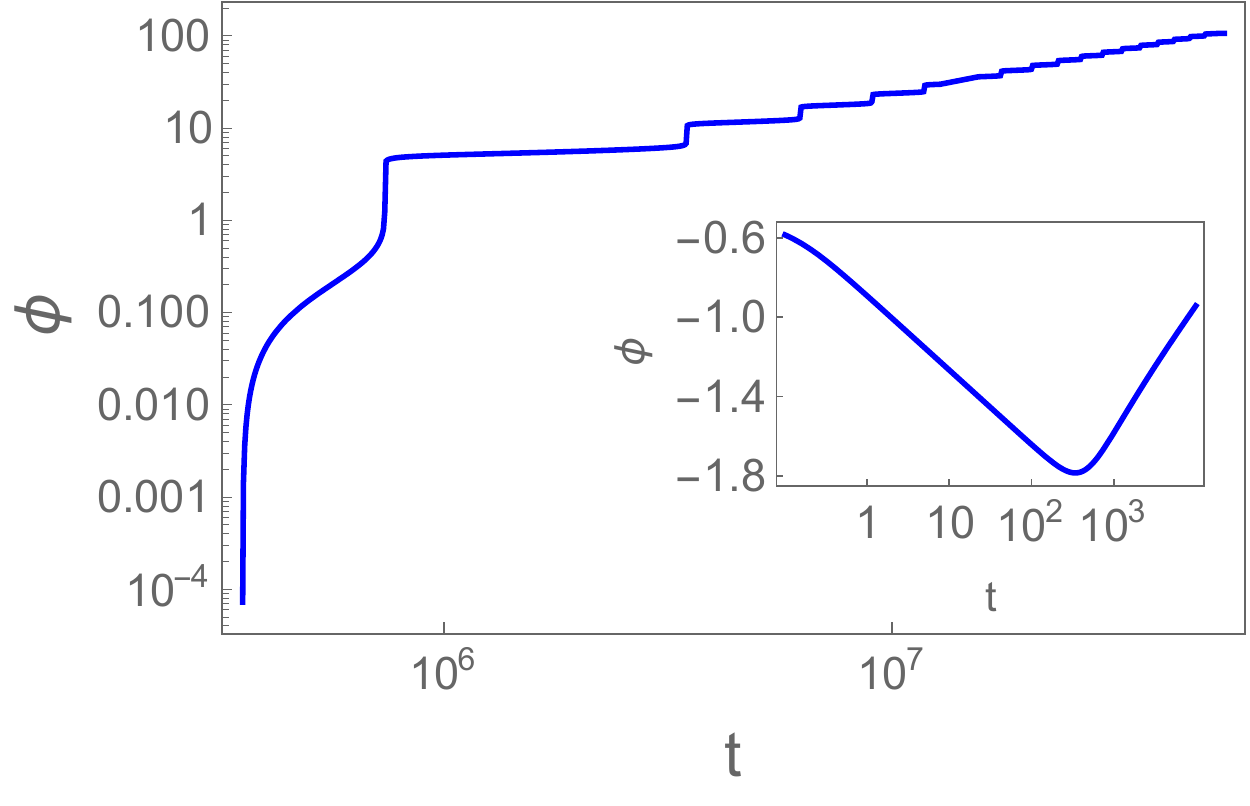}
\caption{This figure shows the volume, equation of state, density, and scalar field as a function of time for the Starobinsky potential with the initial conditions specified by (4.4). The initial conditions are set at the bounce, which takes place at the maximum allowed density. The initial energy density is dominated by the kinetic energy. This universe undergoes many bounces before entering inflation. The inflaton rolls up and down the left wing of the potential, up the right wing, and then inflation occurs as it rolls down the right wing.}
\label{f7}
\end{figure}

The next example concerns Fig. \ref{f7}, which corresponds to the initial conditions
\bq
\lb{initial5}
v_0=3.11\times10^8,\quad \quad \phi_0=-0.50.
\eq
In this case, the initial bounce at $t=0$ is dominated by the kinetic energy of the scalar field. The universe undergoes a longer series of bounces and recollapses before the onset of inflation as is depicted in the top left panel of Fig. \ref{f7}.  With a large initial velocity, the inflaton, which starts from the left wing, gets much higher up on the right wing than in the two previous cases. Hence it takes longer for the inflaton to turn around, leading to many more cycles in the pre-inflationary phase. All the bounces in the pre-inflationary phase happen at the Planck energy density around $\rho_\mathrm{crit}$ as is shown for the first four bounces. Meanwhile, all the recollapses occur at the minimum energy density far below the Planck scale. Similar to the previous two cases, following the turnaround point on the right wing of the potential, inflation takes place at an energy density of around $10^{-13}$ times the Planck density. In addition to the volume of the universe and its energy density, the equation of state, which is shown in the top right panel, also oscillates with the same period as the volume. At each bounce, the equation of state reaches a maximum, while with each consecutive recollapse in the pre-inflationary phase it reaches a lower minimum value, slowly approaching $-1/3$, marked by the upper red dashed line in the plot. We see that even though the  equation of state has a value close to $-1/3$ during the first recollapse, it subsequently takes many cycles for it to cross this value and inflation to begin. This is due to the fact that these initial conditions give rise to a weak hysteresis, which is also evident from the volume plot. The volume at which the universe recollapses increases in consecutive cycles, resulting in a decrease in the minimum energy density at each recollapse. With each cycle the curvature effect is slightly weaker, and hence each cycle brings the universe closer to the right conditions for inflation to begin. Though weaker in strength than in previous cases, the increasing hysteresis prevents the universe from undergoing infinitely many cycles of expansion and contraction, instead facilitating the onset of an inflationary period in a closed LQC universe after a finite number of cycles.

\begin{figure}
\includegraphics[width=220pt]{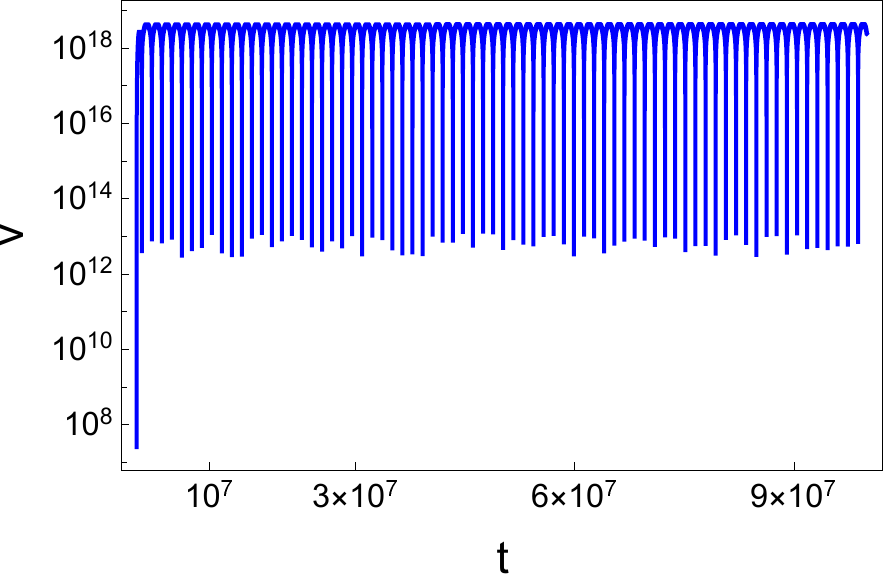}
\includegraphics[width=220pt]{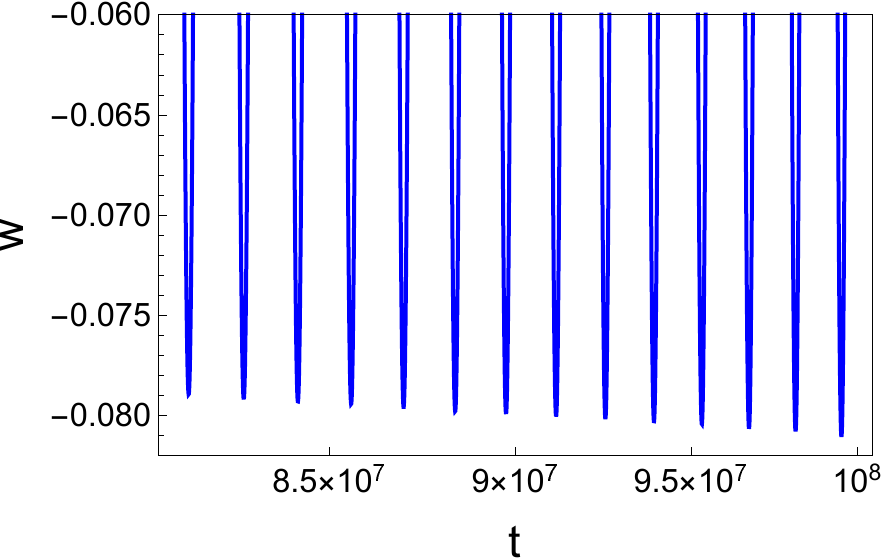}
\includegraphics[width=220pt]{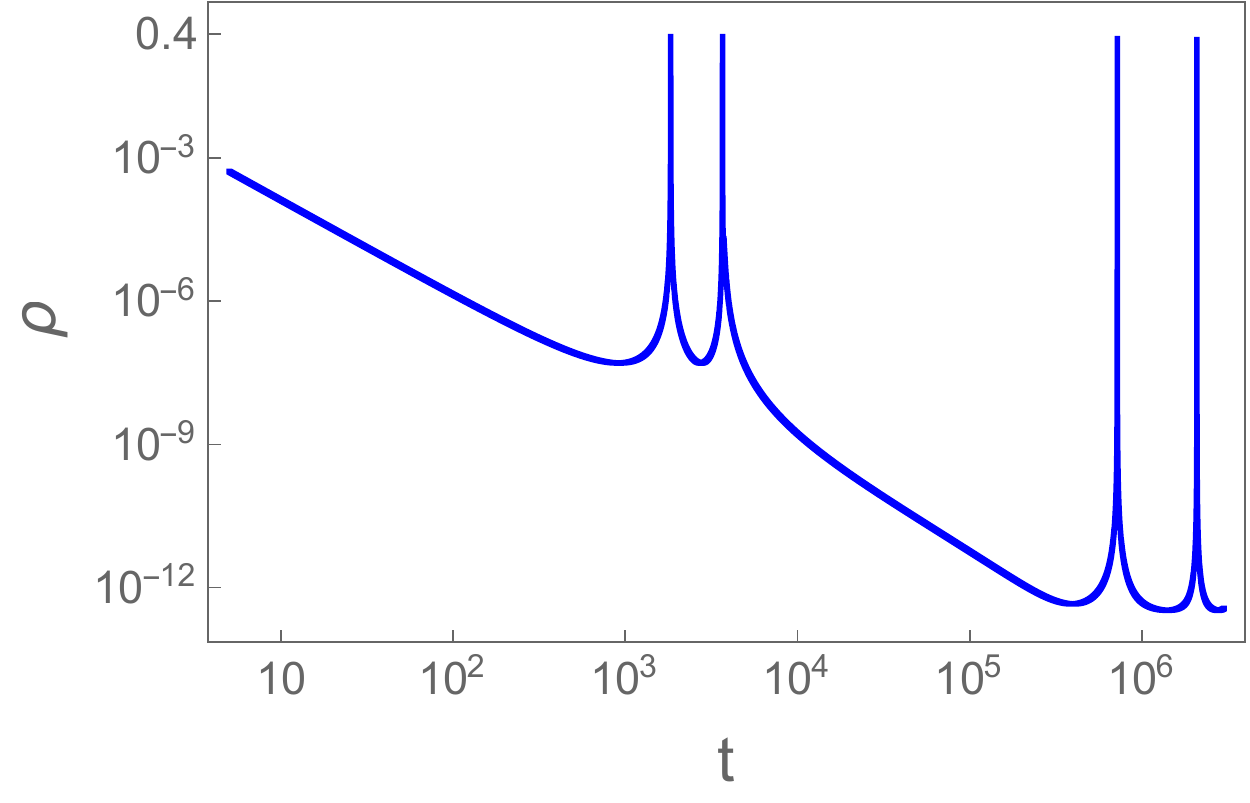}
\includegraphics[width=220pt]{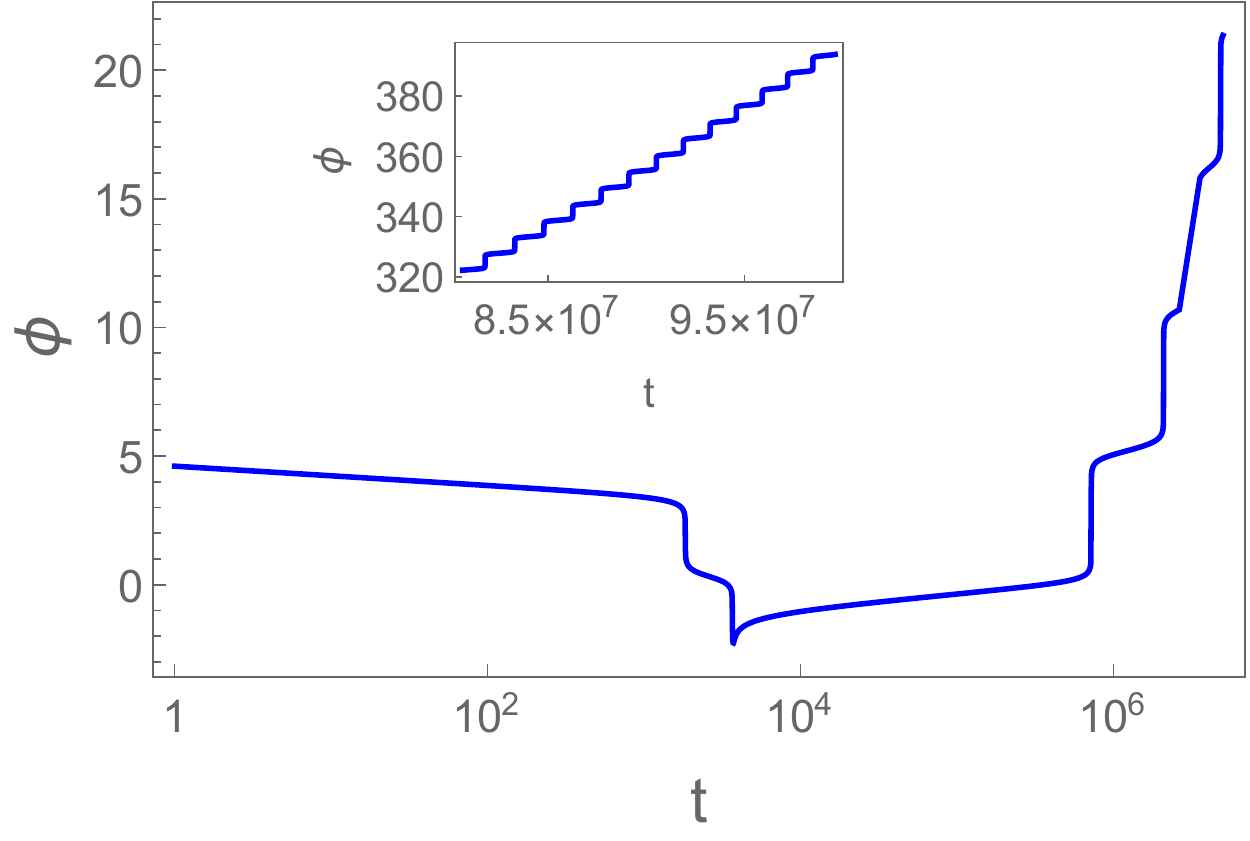}
\caption{This figure shows the volume, equation of state, density, and scalar field as a function of time for the Starobinsky potential with the initial conditions specified by Eq. (\ref{initial6}). The initial energy density at the bounce is dominated by the kinetic energy. This universe undergoes many bounces and we are unable to see the onset of inflation up through the time we have plotted. The inflaton rolls down the right wing of the potential, up and down the left wing, and then rolls up the right wing, not turning around in the time shown here.}
\label{f8}
\end{figure}

Fig. \ref{f8} is an example of a case in which inflation does not appear in our simulation, which ran through time $10^8$. The initial conditions are given by 
\bq
\lb{initial6}
v_0=2.50\times10^7,\quad \quad \phi_0=5.00.
\eq
In this case, the inflaton is released from the right wing of the potential with a large initial velocity, climbs up the left wing of the potential and then turns around, finally climbing up the right wing. The volume plot shows a very large number of bounces and recollapses. Similar to the previous cases, all the bounces occur at the maximum energy density in the Planck regime while all the recollapses take place at the minimum energy density, which is far below the Planck energy. For the time range of the simulation, which is $10^8$ (in Planck seconds), the volume plot shows a long cyclic phase and no inflation. However, as in the previous case where we did observe the onset of inflation, an increasing hysteresis can be seen in the $w$ plot, which shows the minima of the equation of state decrease with each recollapse. The decrease with each cycle is very small, and since towards the end of the evolution shown the minima have only reached -0.08, it is evident that far more time is needed before they can cross -1/3. Note that the decrease in the minimum of $w$ with each cycle was also small in the previous case, but because at the beginning of the oscillatory phase the minima were already close to -1/3, we were able to see inflation before time $10^8$. The minima of $w$ being higher in this case at the start of the oscillatory phase reflects the fact that these initial conditions are less favorable for inflation than in the previous cases, where either the minima of $w$ were close to -1/3 at the start of the oscillatory phase or the initial conditions were such that the hysteresis effect was stronger and so an extremely long series of cycles was not necessary to bring about the onset of inflation.

\subsection{Phase space portraits}

To understand the qualitative dynamics of general solutions with various initial conditions, we present phase space portraits for the Starobinsky potential using 
\bq
\lb{4b1}
X=\chi_0\left(1-e^{-\sqrt{\frac{16\pi G}{3}}\phi}\right), \quad \quad Y=\frac{\dot \phi }{\sqrt{2\rho_{\mathrm{max},0}}},
\eq
where $\chi_0=\sqrt{\frac{3m^2}{32\pi G \rho_{\mathrm{max},0}}}$ and $\rho_{\mathrm{max},0}$ stands for the initial maximum energy density. These variables obey the following set of first-order differential equations 
\bqn
\lb{4b2}
\dot X&=& mY\left(1-X/\chi_0\right),\\
\lb{4b3}
\dot Y&=& -3 HY-mX\left(1-X/\chi_0\right).
\eqn
Together with (\ref{friedmann}), the above  equations (\ref{4b2})-(\ref{4b3}) form a complete set of dynamical equations for a system that is described by $X$, $Y$, and $v$. Since the main goal of our numerical analysis is to determine whether the inflationary phase is a local attractor for a variety of initial conditions, we focus on a 2-dimensional phase space plot in the subspace spanned by $X$ and $Y$. From Eqs. (\ref{4b2})-(\ref{4b3}), it follows that there are two fixed points in the system, which are $(X=0,Y=0)$ and $(X=\chi_0,Y=0)$. The first fixed point corresponds to the end of the reheating phase where the energy density of the inflaton field is very small and the volume of the universe is very large. The second fixed point is located along the $X=\chi_0$ boundary separating the region with real values of the scalar field from the region where the scalar field is complex \cite{lsw2018b}. In the following, we will only focus on the region in which the scalar field is real (to the left of the boundary line). As for the $\phi^2$ potential, in order to make the various solutions converge in a short time, we use a larger value for the mass associated with the scalar field in the phase space portraits. The use of such a value does not change the qualitative behavior of the solutions, including the existence of inflationary separatrices and cosmological attractors, which are the properties of interest. 

\begin{figure}
\includegraphics[width=220pt]{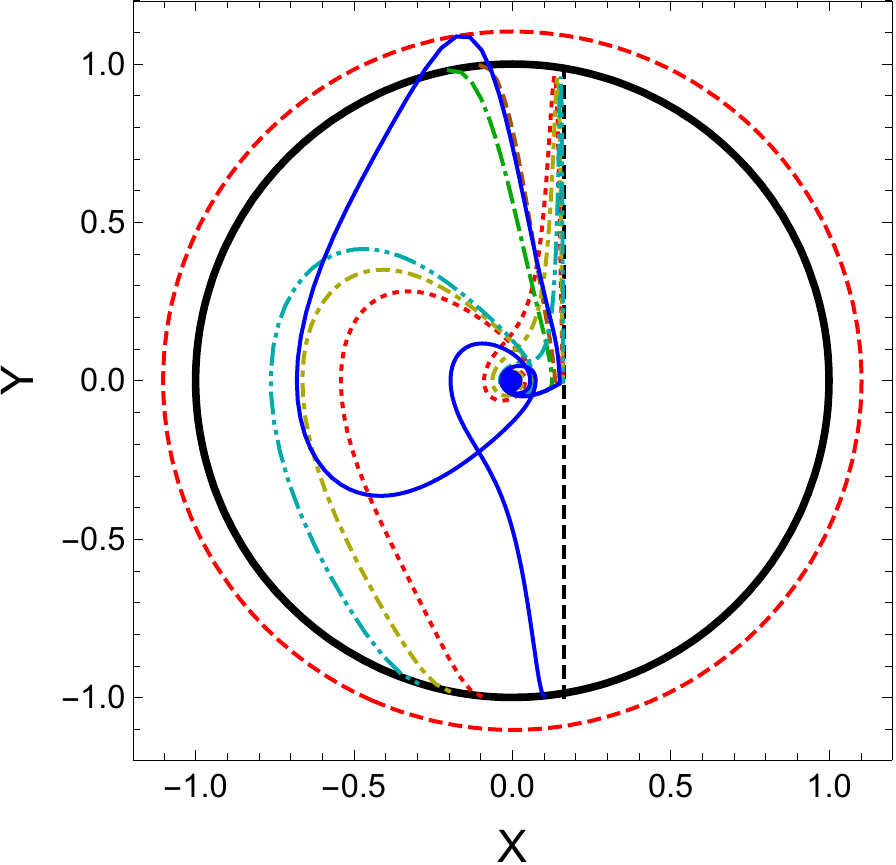}
\includegraphics[width=220pt]{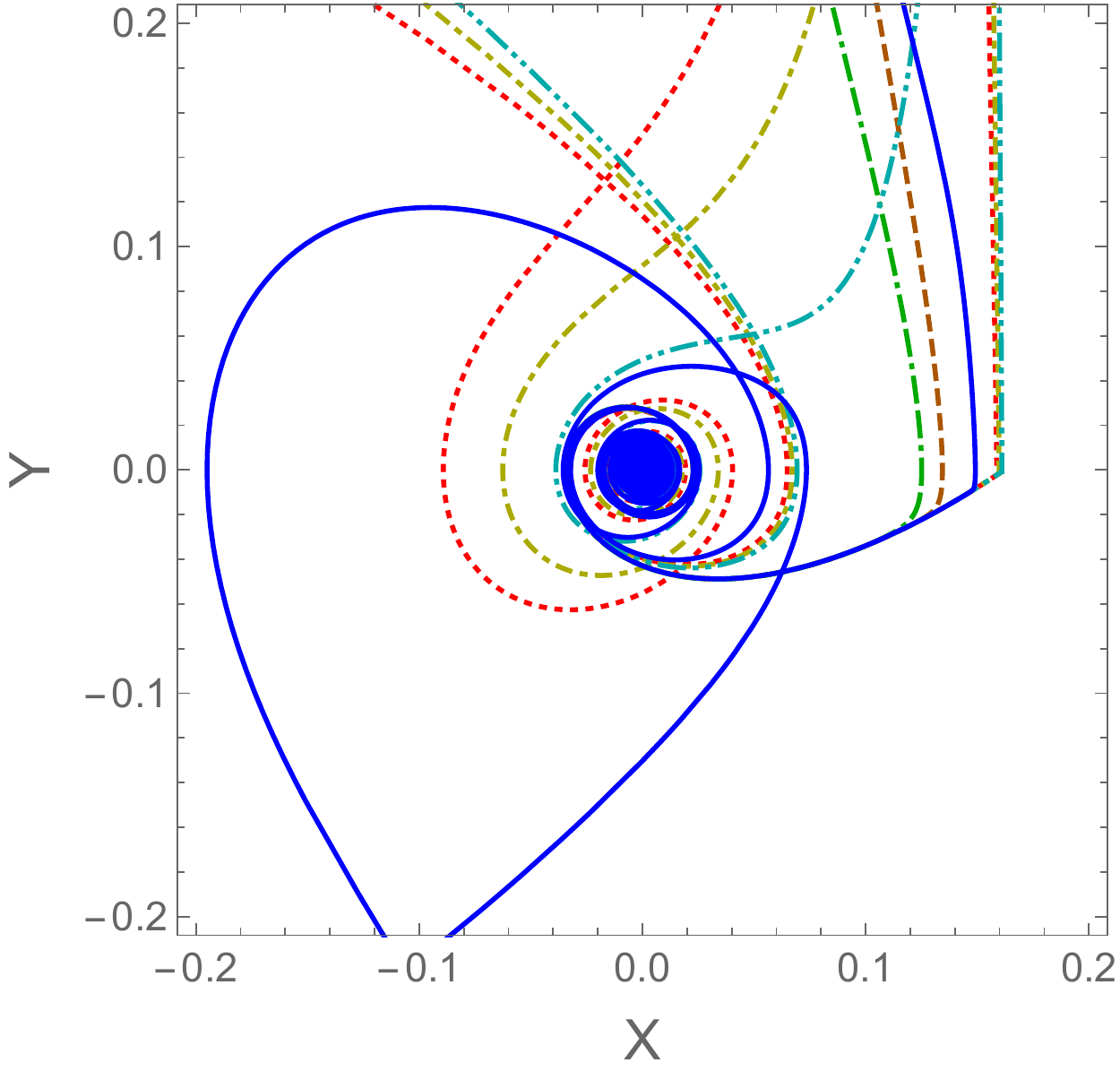}
\includegraphics[width=225pt]{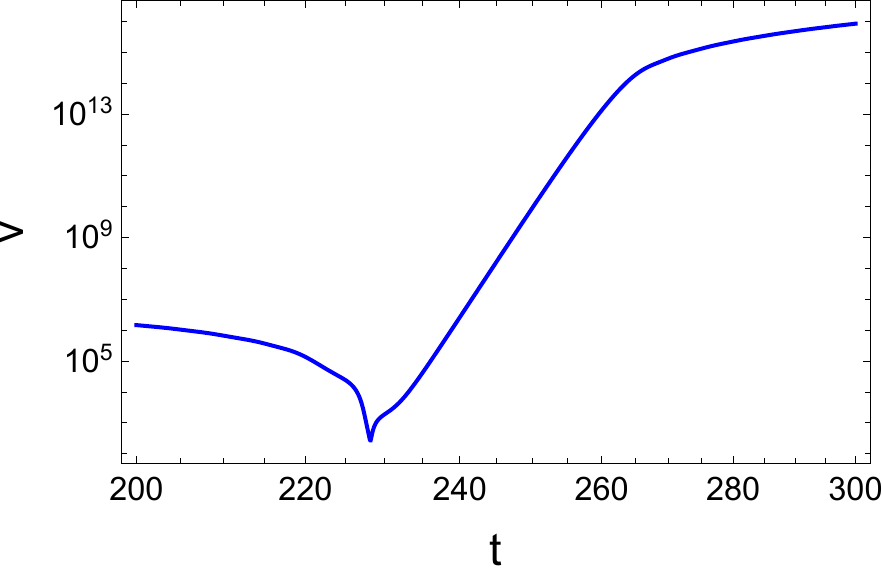}
\includegraphics[width=220pt]{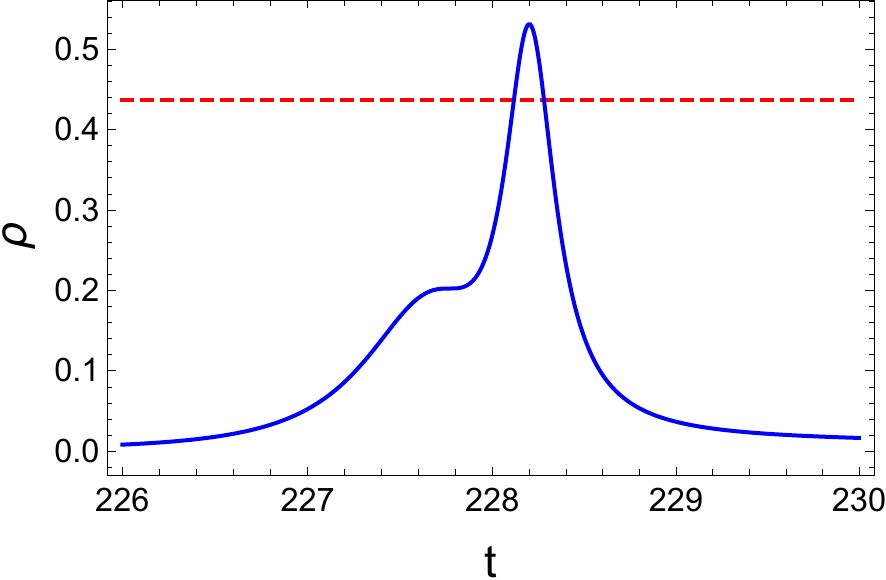}
\caption{Phase space portrait, volume, and density plots for the Starobinsky potential with $m=0.62$ and initial volume $v_0=1000$. There are six trajectories shown, each corresponding to a distinct set of initial conditions. For all the curves, the initial maximum density is $\rho_{\text{max},0}=0.44$ and $\chi_0=0.16$. The red dashed circle in the phase space portrait represents the maximum density achieved during any of the cycles shown. The red dashed line in the density plot represents the maximum density at the initial time.} 
\label{f9}
\end{figure}

The first phase space portrait is presented in Fig. \ref{f9}. The upper left plot shows the entire phase space region which includes trajectories for six distinct initial conditions. For the Starobinsky potential only the trajectories confined to the left of the vertical black line $X=\chi_0$ correspond to real values of the  scalar field. Thus, these are the only cases we consider in the phase space plots. The phase space evolution of a generic solution with one bounce is represented by the solid blue curve in the phase space portrait, which corresponds to  the  initial conditions $(\phi_0 = 0.23,  p_{\phi_0}=-930)$ at $v_0=100$.
The trajectory starts from the black solid circle which corresponds to the initial bounce at the maximum energy density. It then moves towards the origin as the scalar field loses energy. When the blue curve is close to the origin, a recollapse occurs and the universe enters into a contracting phase, increasing the energy density of the scalar field and making the blue curve move away from the origin. Afterwards, as the energy density of the scalar field increases, a bounce takes place when the blue curve hits the dashed red circle and the maximum energy density at that time is reached. After the bounce, the universe re-enters a state of expansion and the scalar field starts to lose energy again. The blue curve then moves towards the origin and merges into the inflationary separatrix (the short, curved, horizontal line traveling left towards the origin, during which inflation takes place) before finally falling into the spiral at the center of the plot. The other trajectories have similar qualitative behavior to the blue solid curve. The difference lies in the number of bounces in the pre-inflationary phase, which for this plot is either none or one, as well as the value of the maximum energy density at which the bounce occurs. For example, the dot dashed green curve starts from the top and moves directly towards the inflationary separatrix without any further bounces, while the red dotted, yellow dot-dashed, and green dot-dot-dashed curves all experience a single bounce that happens at the highest allowed energy density, which is close to the initial energy density. Compared with $\phi^2$ inflation, where we saw relatively long and straight horizontal lines heading towards the origin, the inflationary separatrix in Starobinsky inflation is significantly less noticeable. For this reason, we show the upper right panel which zooms in on the inflationary separatrix, where all the curves merge, and the attractor at the origin.
Finally, to show the details of the evolution of a generic solution in the phase space portrait, the bottom panels show the behavior of the volume and energy density for the solid blue curve. We see that in the volume plot on the bottom left, the single bounce happens at around $t=228$, while in the density plot on the bottom right we see that the energy density at this bounce exceeds the initial energy density, represented by the dashed red line in the graph. 

\begin{figure}
\includegraphics[width=220pt]{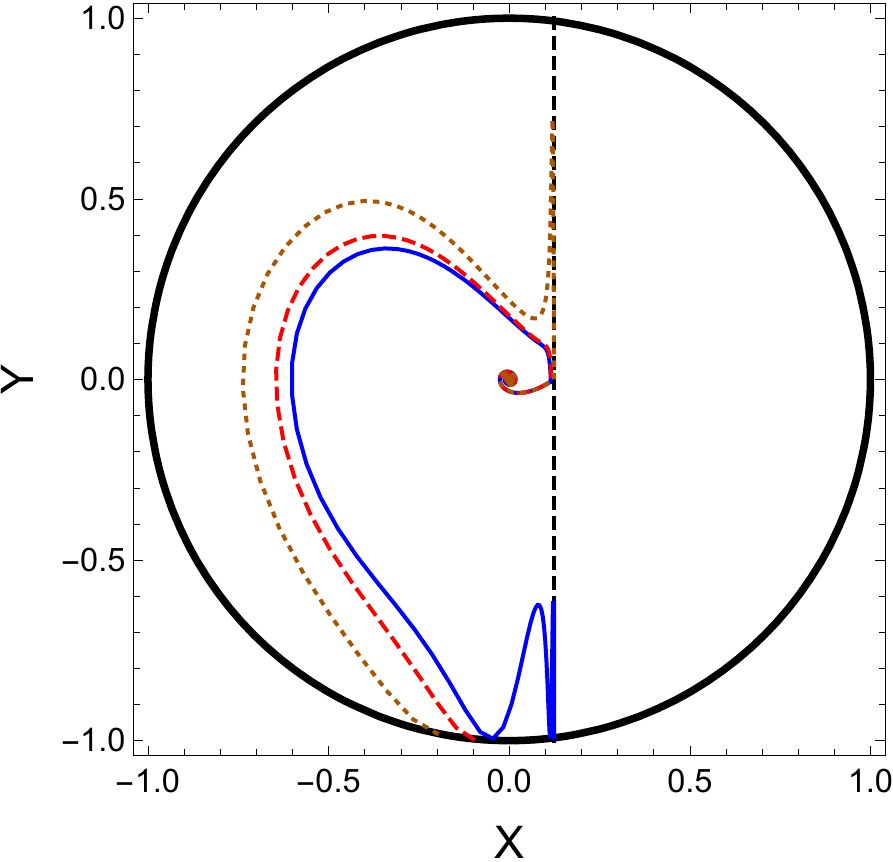}
\includegraphics[width=220pt]{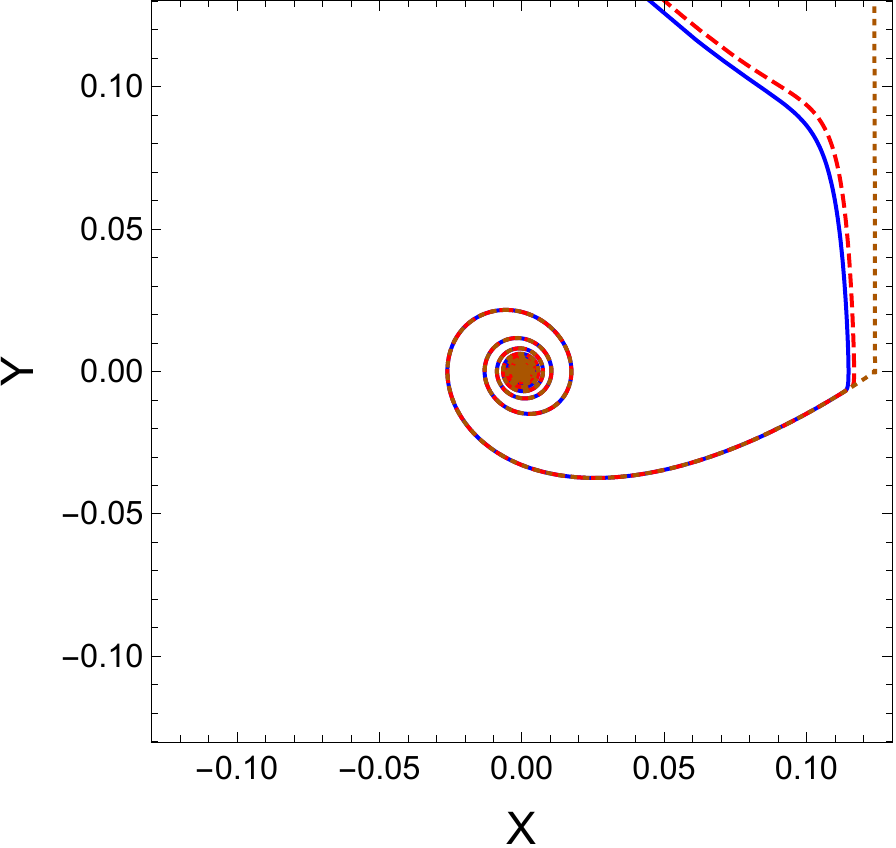}
\includegraphics[width=220pt]{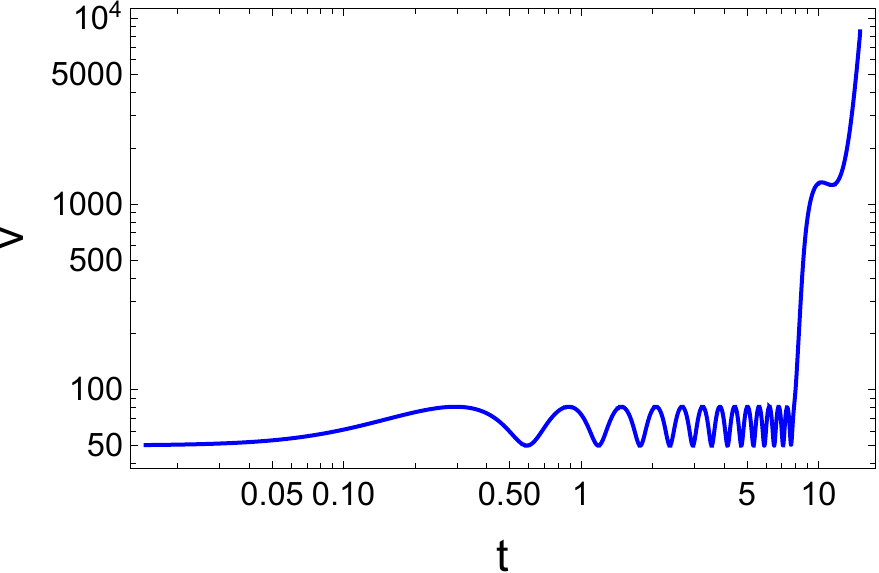}
\includegraphics[width=220pt]{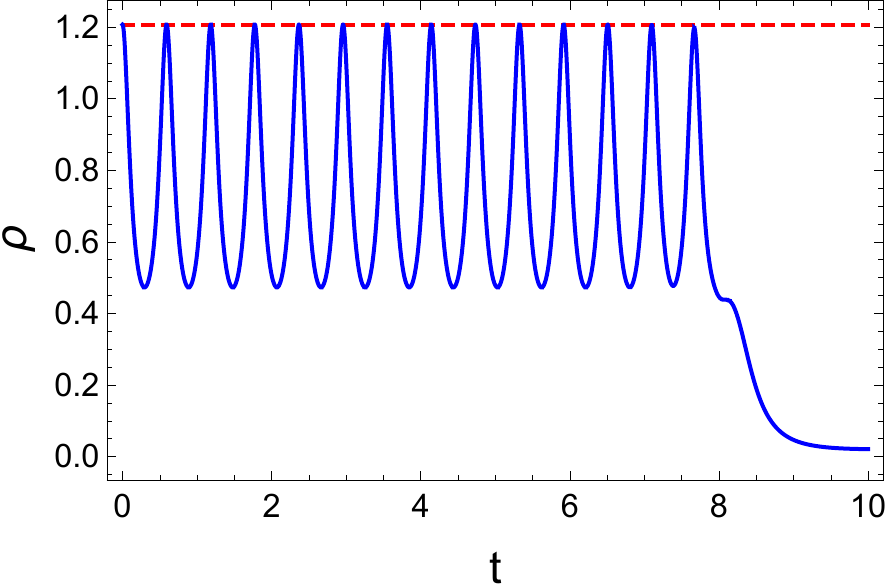}
\caption{Phase space portrait along with volume and density plots for the Starobinsky potential with $m=0.79$ and initial volume $v_0=50$. For all the curves, the initial maximum density is $\rho_{\text{max},0}=1.21$ and $\chi_0=0.12$. 
The red dashed line in the density plot shows the maximum density at the initial time.} 
\label{f10}
\end{figure}

The second phase space portrait presented in Fig. \ref{f10} includes trajectories corresponding to three distinct initial conditions. All of them describe universes that undergo bounces before inflation. The upper left plot shows the entire phase space region. For the universes corresponding to these initial conditions, the energy density never exceeds the initial density. As a result, all the trajectories are confined within the initial black unit circle. The representative solution is again displayed as the solid blue curve, which is analyzed further in the bottom panels. The initial conditions for the blue curve are $(\phi_0, p_ {\phi_0})=\left(8.97,-77\right)$ at $v_0=50$. As shown in the bottom panels, the universe starts at a bounce and then undergoes a series of cycles with alternating contracting and expanding phases until inflation begins. All the bounces happen at the maximum energy density with a volume no larger than the initial volume, while all the recollapses happen at the minimum energy density for each cycle, all of which are also in the Planck regime as can be seen in the $\rho$ plot. In the top right panel, we zoom in on the inflationary separatrix and the attractor at the center. Since all the recollapses happen in the Planck regime, there are no curves that approach the origin before merging into the separatrix. We can clearly see from the plot that trajectories starting from different initial conditions have the same late-time dynamics consisting of an inflationary phase followed by a reheating phase.

\section{Summary}
\label{summary}
\renewcommand{\theequation}{4.\arabic{equation}}\setcounter{equation}{0}

The goal of this manuscript is to understand the onset of inflation in closed universes for low energy scale inflationary models, such as Starobinsky inflation. Starting inflation in such cases has remained a long-standing problem because of the recollapse caused by the spatial curvature and the big crunch singularity that are unavoidable in the classical theory\color{black} \cite{lindelecture,linde2018}. We explored a resolution of this problem in the setting of LQC where big bang/big crunch singularities are robustly resolved due to non-perturbative quantum gravity effects and the pre-inflationary phase is in general characterized by a series of bounces and recollapses of the universe.
For comparison we first considered the case of $\phi^2$ inflation and demonstrated that non-singular cyclic evolution in the pre-inflationary phase sets the stage for inflation to begin even for very unfavorable initial conditions. The analysis was then repeated for Starobinsky inflation, where the problem is far more severe, yielding similar results. 

 For the $\phi^2$ potential, inflation can take place at different energy scales ranging from the Planck regime to an energy density that is $10^{12}$ orders of magnitude below the Planck density. Inflation can also take place on both sides of the potential, resulting in two inflationary separatrices in the phase space portraits. On the other hand, for the Starobinsky potential, with the mass parameter fixed by observations, inflation can only take place on the right wing of the potential at an energy density that is around $10^{13}$ orders of magnitude below the Planck density. Due to the asymmetry of the potential, there is only one inflationary separatrix in the phase space portraits. We found different features in terms of the pre-inflationary dynamics, which is when the cycles take place. In the $\phi^2$ model, both the bounces as well as the recollapses can occur at the maximum energy density, which means that $t=0$ where the energy density takes on the maximum allowed value can correspond to either a bounce or  a recollapse. In the Starobinsky model, on the other hand, the bounces always happen at the maximum energy density and the recollapses happen at the minimum energy density, which means that $t=0$ always corresponds to a bounce. Furthermore, for the Starobinsky potential it tends to take longer for the first recollapse to occur, and there also tends to be more time between subsequent cycles than for the $\phi^2$ potential. This contributes to the delay in the onset of inflation in the Starobinsky model as compared with the $\phi^2$ model. With respect to which initial conditions give rise to inflation, we found that the evolution of the universes in both models differ most in those cases where the oscillatory behavior of $w$ begins with the minima of $w$ not close to $-1/3$. In such cases, with each cycle the minimum of $w$ decreases noticeably for the $\phi^2$ potential, so that after some relatively small, finite number of cycles $w$ crosses -1/3 and inflation begins. For the Starobinsky potential, however, the minima of $w$ decrease very slowly with each cycle. This means that when starting with initial conditions for which the minima of $w$ are not close to -1/3 when the oscillatory behavior in pre-inflationary epoch begins, then it can take an extremely long time for the minima of $w$ to decrease enough to cross -1/3. Thus, in comparison to $\phi^2$ potential, the mechanism resulting from non-singular cyclic evolution and leading to the onset of inflation in the Starobinsky potential is weaker but nevertheless strong enough to overcome problems encountered in the classical theory.

In summary, we have shown that the problem of the onset of inflation for low energy scale models, such as the Starobinsky potential, in a closed universe can be successfully resolved by quantum gravity effects. The primary reason for the initiation of the inflationary phase is a progressive decrease in the value of  the  equation of state with each cycle captured in the hysteresis-like phenomena seen earlier for the $\phi^2$ potential in LQC \cite{ds2020} and other bouncing models \cite{st2012,sst2015}. When the equation of state becomes less than $-1/3$ in a particular cycle, recollapses no longer occur and inflation starts. While for the $\phi^2$ potential we found inflation to occur in all the considered cases, for the Starobinsky inflation we found some cases in which the hysteresis-like phenomenon is so weak  that the onset of inflation, though expected, is delayed. It would be interesting to understand  physical phenomena that can make  the onset of inflation even in such extreme cases more favorable, an example of which will be discussed in a future work \cite{ms}. Apart from such cases, thanks to the singularity resolution due to non-perturbative quantum gravity effects, the Starobinsky potential results in inflation in a short time even  when starting from initial conditions which are highly unfavorable for inflation in the classical theory.

\section*{Acknowledgements} 
This work is supported by the NSF grants PHY-1454832 and NSF PHY-1852356. L.G. thanks the REU program of the Department of Physics and Astronomy at LSU during which most of this work was carried out.

\end{document}